\documentclass[superscriptaddress,10pt,aps,pra,twocolumn,longbibliography, notitlepage]{revtex4-1}
\usepackage{float}
\floatstyle{boxed}
\usepackage{array}
\usepackage{graphicx}
\usepackage{amsbsy}
\usepackage[utf8x]{inputenc}
\usepackage{epstopdf}
\usepackage{amsmath,amssymb,amsfonts,amsthm}
\usepackage{indentfirst}
\usepackage{soul}
\usepackage[T1]{fontenc}
\usepackage[dvipsnames]{xcolor}
\usepackage{url}
\usepackage[colorlinks]{hyperref}

\usepackage{stackengine}
\stackMath
\hypersetup{
    plainpages=true,
    breaklinks=true,
    hypertexnames=false,
    pageanchor=true,
    colorlinks=true,
    linkcolor={blue},
    citecolor={red},
    urlcolor={blue},
    anchorcolor={black}
}

\newcommand{\figref}[1]{\mbox{Fig.~\ref{#1}}}

\renewcommand{\eqref}[1]{\mbox{Eq.~(\ref{#1})}}

\newcommand{\be}{\begin{equation}}
\newcommand{\ee}{\end{equation}}
\newcommand{\bea}{\begin{eqnarray}}
\newcommand{\eea}{\end{eqnarray}}

\begin{document}

\title{How to realize compact and non-compact localized states \\ in disorder-free hypercube networks}
\author{Ievgen I. Arkhipov}
\email{ievgen.arkhipov@upol.cz}
\affiliation{Joint Laboratory of
Optics of Palack\'y University and Institute of Physics of CAS,
Faculty of Science, Palack\'y University, 17. listopadu 12, 771 46
Olomouc, Czech Republic}

\author{Fabrizio Minganti}
\thanks{Present address: Alice \& Bob, 53 boulevard du Général Martial Valin, 75015 Paris, France}
\affiliation{Laboratory of Theoretical Physics of Nanosystems (LTPN), Institute of Physics, \'{E}cole Polytechnique F\'{e}d\'{e}rale de Lausanne (EPFL), 1015 Lausanne, Switzerland}
\affiliation{Center for Quantum Science and Engineering, \\ \'{E}cole Polytechnique F\'{e}d\'{e}rale de Lausanne (EPFL), CH-1015 Lausanne, Switzerland}

\author{Franco Nori}
\affiliation{Quantum Information Physics
Theory Research Team, Quantum Computing Center, RIKEN, Wakoshi,
Saitama, 351-0198, Japan} \affiliation{Physics Department, The
University of Michigan, Ann Arbor, Michigan 48109-1040, USA}

\begin{abstract}
We present a method for realizing various zero-energy localized states on disorder-free hypercube graphs. Previous works have already indicated that disorder is not essential for observing localization phenomena in noninteracting systems, with some prominent examples including the 1D Aubry-Andr\'e model, characterized solely by incommensurate potentials, or 2D incommensurate Moir\'e lattices, which exhibit localization  due to the flat band spectrum.  Moreover, flat band systems with translational invariance can also possess so-called compact localized states, characterized by exactly zero amplitude outside a finite region of the lattice. 
Here, we demonstrate that both compact and non-compact (i.e., Anderson-like) localized states {naturally} emerge in disorder-free hypercubes, which can be systematically constructed using Cartan products. {This construction ensures the robustness of these localized states against perturbations. Furthermore, we show that the  hypercubes can be associated with the Fock space of interacting spin systems exhibiting localization.
Viewing localization from the hypercube perspective, with its  inherently simple eigenspace structure, offers a clearer and more intuitive understanding of the underlying Fock-space many-body localization phenomena.
Our findings can be readily tested on existing experimental platforms, where hypercube graphs can be emulated, e.g., by photonic networks of coupled optical cavities or waveguides. The results can pave the way for the development of novel quantum information protocols and enable effective simulation of quantum many-body localization phenomena.} 
\end{abstract}

\date{\today}

\maketitle

\section{Introduction}
The disorder-induced localization phenomenon has been known since the seminal work of P. W. Anderson studying the quantum phase transition between an isolating and metallic phase~\cite{Anderson1958}.
In the presence of disorder, the wavefunctions of electrons confined by one- and two-dimensional periodic potentials become exponentially localized in space, as multiple scattering processes generate destructive interference between the otherwise delocalized modes.
In higher-dimensional disordered systems $D\geq3$, there can coexist both extended and localized states, and the energy level, separated these two, is known as the mobility edge~\cite{Mott1987,Lagendijk2009,AbrahamsBook,Bliokh2012,LifshitsBook}. 

{However, later it was realized that a disorder is not an essential ingredient for observing localization phenomena.
For instance, a periodically kicked quantum rotator can be related to the Anderson localization problem in  one-dimensional disordered lattices~\cite{Fishman1982}.}
Localization can also manifest in quasiperiodic systems without disorder, such as the 1D Aubry-André model, which is characterized by a quasiperiodic, i.e., incommensurate, potential energy~\cite{Aubry1980,Grempel1982}. Incommensurate potentials in a Hamiltonian have a quasiperiodic modulation that does not align rationally with the lattice. In 2D, the localization transition can occur in Moir\'e lattices~\cite{Wang2019_nat}, Vogel spirals~\cite{Sprignuoli2019}, and in linear~\cite{Zhu2024} or nonlinear quasicrystals~\cite{Freedman2006}.
Moreover, even in purely periodic, i.e., translational invariant, systems without quenched disorder, localization can arise due to many-body interactions.
A primary example is the Mott transition from insulator to metal (superfluid) in fermionic (bosonic) systems, both demonstrated using ultracold gasses quantum simulators \cite{Greiner2002,Jordens2008}, where the localization properties are rather triggered by the presence of Coulomb-like interactions.
Furthermore, in 1D lattice gauge theories, localization can result from gauge superselection sectors that act as an effective internal quenched disorder~\cite{Smith2017,Brenes2018}. In 2D periodic systems, localization may occur due to emergent classical percolation transition that divides the system into isolated real-space clusters~\cite{Chakraborty2022}.
A form of localization can manifest also in non-interacting and periodic systems through the so-called compact localized states (CLSs), i.e., wavefunctions whose amplitude strictly vanishes outside a finite domain of a system~\cite{Vicencio2021,Danieli2024}. These CLSs emerge from destructive interference of macroscopically degenerate eigenstates whose existence is determined by a flat-band energy spectrum~\cite{Nori1990,Flach2014,Maimaiti2017,Rontgen2018let,Rontgen2018,Leykam2018,Leykam2018b,Danieli2020}.  

Apart from Hermitian systems, the localization phenomena can also be observed in disordered non-Hermitian systems as well~\cite{Hatano1996,Hatano1997}.
Additionally, it was found that the bulk-boundary correspondence failure in non-Hermitian systems is intrinsically related to the so-called non-Hermitian skin effect where a number of edge modes exponentially become localized at the boundaries~\cite{Yao2018,Alvarez2018,arkhipov2023a,Ozawa2019,Bergholtz2021}.   
These theoretical findings have been further experimentally validated in photonic platforms~\cite{Weimann2017,Roy2021b,Weidemann2020}.

Here we investigate emerging and {\it controllable localization occurring in disorder-free systems} whose geometry is that of a hypercube graph.
Similar hypercube structures have been studied in connection with various phenomena. In the classical realm, they are closely related to the mutation-selection models of population genetics, which seeks to predict gene susceptibility or resistance to mutations~\cite{Konig206}. In the quantum field, hypercube structures and their spectral characteristics can be useful for evaluating geometric entanglement in multipartite states~\cite{Semenov2024}, for exhibiting phase-space features significantly smaller than Planck’s constant~\cite{Howard2019}, and even for implementing high-performance fault-tolerant quantum computing~\cite{Goto2024}. Hypercube geometry has also gained interest in condensed matter physics, particularly in the description of spin-glass models~\cite{Parisi1994, Marinari1995}. Additionally, the spectral properties of chaotic hypercube lattices can bear resemblance to the Maldacena-Qi model, which describes wormholes~\cite{Jia2020}.  
Furthermore, hypercubes with {\it{disordered}} potentials have been recently explored, for example, in the context of continuous parabolic Anderson models~\cite{Avena2020}, discrete models with quantum walks~\cite{Stefanak2023}, {and Fock-space many-body localization~\cite{Laumann2014,Baldwin2016,Roy2020mbl,Logan2019,Scoquart2024}.}

Specifically, we demonstrate that both zero-energy CLSs and non-compact localized states (NCLSs) can be {realized in disorder-free hypercube networks. These graphs can be readily emulated, e.g., by a bosonic network of coupled cavities and waveguides, or implemented in high-dimensional photonic synthetic spaces}~\cite{Regensburger2011,Yuan2018,Hu20,Leefmans2022,Parto2023,Ehrhardt2023,Leefmans2024}.  
The key difference between the two is that while CLSs have strictly zero amplitudes beyond a finite region in the lattice, NCLSs do not, making the latter more akin to Anderson localized states,  {exhibiting exponentially decaying site populations around a few pronounced eigenstate intensity peaks.}

We show that in the case of identical site potentials, the hypercube spectrum exhibits macroscopic degenerate states, similar to those found in flat-band systems, whose destructive interference results in CLSs. 
Conversely, (in)commensurate potentials without disorder can produce NCLSs
with a controlled (single-site) periodic amplitude density. {These findings are in contrast with previous studies~\cite{Avena2020,Laumann2014,Baldwin2016,Roy2020mbl,Scoquart2024}, which exclusively attribute hypercube localization, when mapped to the Fock space of spin systems,  to disorder.}

We describe a constructive procedure to obtain the parameters needed to generate these states, based on a recursive application of Cartan products to the basic building blocks of the hypercubes known as dions. This construction ensures that the engineered localized ZESs of the hypercube are robust against various perturbations and disorder. {We additionally reveal that specifically weighted hypercube graphs can be associated with the Fock space of interacting spin-1/2 systems, with or without disorder, providing thus deeper insights into the origin and existence of robust many-body localized states in such systems~\cite{Logan2019}. In this respect, we note that despite the existent literature on Fock-space many-body localization in spin systems~\cite{Laumann2014,Baldwin2016,Roy2020mbl,Logan2019,Danieli2020}, its explicit connection with hypercube space is often overlooked. However, viewing localization from the hypercube perspective, with its {\it inherently simple eigenspace structure}, offers a clearer and more intuitive understanding of the underlying phenomena.}   

{Our results further suggest that linear hypercube networks can provide a promising practical platform for implementation of various quantum information protocols, particularly for quantum storage~\cite{Rontgen2018let}, and the simulation of both the flat-band and disorder-induced-like many-body localization~\cite{Vicencio2021,Danieli2024,Logan2019}. This could pave the way for advancements in quantum information processing and effective simulation of various quantum many-body models.}

This paper is structured as follows: In Sec.~\ref{I}, we give a brief summary of our main results on localization on hypercube graphs characterized by ordered site potentials. In Sec.~\ref{II}, we introduce and describe a general method for constructing hypercubes with certain ordered site potentials and outline its main properties. Section~\ref{III} focuses on the applicability of this method for engineering CLSs on the example of 8D hypercubes. {There, we also explain the similarities between the CLSs in the hypercube networks and many-body flat-band localization encountered in interacting spin-1/2 systems.} In Sec.~\ref{IV}, we explore the construction of NCLSs with both single-site and periodic amplitude densities and their robustness against imposed correlated and uncorrelated disorder. {The similarity between NCLSs on such perturbed hypercube networks and Fock-space many-body localization is discussed in Sec.~\ref{V}.} The conclusions and outlook are provided in Sec.~\ref{VI}.

\section{Overview of the main results}\label{I}
In this work we focus on the study of the eigenvalue problem of a Hamiltonian
\begin{equation}
    H\psi=E\psi,
\end{equation}
which  describes a certain {\it disorder-free} hypercube graph.
Such a graph can emulate, e.g., a set of coupled waveguides or cavities.  In that case the Hamiltonian $H$ can be written in the mode representation, i.e., $\hat H=\hat\Psi^{\dagger}H\hat\Psi$, where $\hat\Psi=[\hat a_1,\dots,\hat a_n]^T$ is the vector of the bosonic annihilation operators, where an operator $\hat a_j$ represents a mode $j$. The bosonic operators also obey the known commutation relations, namely, $[\hat a_j,\hat a_k]=0$, and $[\hat a_j^{\dagger},\hat a_k]=\delta_{jk}$, with $\delta_{jk}$ being a Kronecker delta function.

More specifically, a bosonic Hamiltonian defined on an $n$-dimensional hypercube graph  can read as
\begin{eqnarray}\label{Hb}
    \hat H = \sum\limits_{i=1}^{2^n}\nu(i)\hat a^{\dagger}_i\hat a_i+g\left(\sum\limits_{j,k\in {\cal N}_j}\hat a_j\hat a^{\dagger}_k+{\rm h.c.}\right),
\end{eqnarray}
where $\nu(i)$ and $g$ account for the potential energy (frequency) at the hypercube site $i$, and the nearest-neighbor site interaction energy, respectively, where ${\cal {N}}_j$ is the set of $n$ vertices constituting the nearest-neighbors of the site $j$.  The potential energy here thus can be considered as the frequency of the mode $\hat a_i$, and the parameter $g$ as the intermode coupling strength. 

{Evidently, the single-particle Hamiltonian in \eqref{Hb} can also describe fermions. However, for concreteness, here we assume that it is of bosonic nature.}

Our main result is that localization phenomena on such a bosonic hypercube graph %, {with, in general, varied intersite (intermode) coupling $g$,} 
can be observed and even controlled when the site potential attains the following ordered form: 
\begin{equation}\label{nu-intro}
    \nu(i)=\sum\limits_{j=1}^{n}\alpha_j {\cal H}(f_{i,j})+\beta_j {\cal H}(-f_{i,j}),
\end{equation}
where $f_{i,j}=\sin\Big[2^j\pi i/(N-1)\Big]$, and ${\cal H}(x)$ denotes the Heaviside step function, i.e.,
\begin{equation}
    {\cal H}(x) = \begin{cases}
        1, \quad x\geq0, \\
        0, \quad x<0,
    \end{cases}
\end{equation}
and $\alpha_j$, $\beta_j$ are, in general, certain real-valued coefficients dependent on the index $j$. 
Depending on the values of $\alpha_j,\beta_j$ the potential energy in \eqref{nu-intro} can exhibit either commensurate or incommensurate behavior. {Incommensurate potentials in a system Hamiltonian refer to spatially varying potentials whose periodicity is incommensurate (i.e., not a rational multiple) with the underlying lattice structure. In contrast, commensurate potentials have periodicities that align rationally with the lattice, typically preserving translational symmetry.}

We explicitly show that while the commensurate case leads to the emergence of compact localized states, the incommensurate case can result in the appearance of non-compact localized states in the hypercube eigenspectrum.
{By demonstrating this, we reveal that localization in a hypercube graph can occur even in the absence of disorder, which contrasts with previous common assumptions~\cite{Avena2020,Laumann2014,Baldwin2016,Roy2020mbl,Scoquart2024}.}

{Another important outcome of our results is that hypercube Hamiltonians with specifically varied mode-coupling strength $g$ in \eqref{Hb}, can be mapped to the Fock space of interacting spin-1/2 systems exhibiting many-body localization. This suggests that linear hypercube networks could also serve as an experimental platform for simulating many-body localization phenomena occurring in the interacting spin systems.}

\section{Theory}\label{II}
\subsection{Hypercube construction}
We first set the stage by describing a general framework for hypercube construction and its spectral characteristics.

Geometrically, an $n$-dimensional hypercube can be readily constructed by iteratively applying the Cartesian product of $n$ 1-dimensional edges (dions)~\cite{CoxeterBook,arkhipov2023c}. 
For instance, in 2D a Cartesian product of two dions generates a square; in 3D, three dions give a cube, and so on.  The resulting $n$-cube has in total $2^n$ vertices and $2^{n-1}n$ edges. 
In a quantum mechanical formalism, the iterative Cartesian products of dions correspond to iterative Kronecker sums applied to the Hamiltonians that describe the dions. Specifically, we choose the description where each dion can be associated with a $2\times 2$ matrix 
\begin{equation}\label{Sk}
S_k=\begin{pmatrix}
            \alpha_{k} & \kappa_k\\
            \kappa_k & \beta_{k},
        \end{pmatrix}
\end{equation}
representing the Hamiltonian of a qubit or that of a linear two-mode system.
Following this choice, a $2^n\times2^n$ Hamiltonian matrix $H_n$ of weighted $n$-hypercube graph can be iteratively constructed. One fixes
\begin{equation} H_1 = S_1
\end{equation}
and then
\begin{eqnarray}\label{An}
    H_n=
        H_{n-1}\otimes I_2+I_{n-1}\otimes
        S_n, \quad n>1,
\end{eqnarray}
where $I_k$ is the identity on the $k$-dimensional Hilbert space.
Importantly, when $\alpha_k=\beta_k=0$, and $\kappa_k=1$, the Hamiltonian takes the form of an ordinary adjacency matrix of a regular $n$-hypercube with zero-potential vertices~\cite{Harary1988}. Throughout the text, without loss of generality, we assume $\kappa_k=1 \, \, \forall k$. {Indeed, since we focus solely on localization in the hypercube eigenspace, its qualitative nature remains unchanged when an arbitrary $\kappa_k$ is absorbed by the diagonal elements of the matrix $S_k$, i.e., when rescaling $\alpha_k$ and $\beta_k$ by $\kappa_k$.} 

In this regard, we note that previous studies on hypercube localization primarily began with hypercubes featuring zero-site potentials, where the diagonal elements were later perturbed in a {\it disordered} manner (see, e.g., Ref.~\cite{Avena2020}). In contrast, here we demonstrate that a more general construction of a weighted hypercube is sufficient to observe localization.

From \eqref{An} it directly follows that $N=2^n$ vertices, or sites, of the $n$-dimensional hypercube with indices $i=0,\dots,N-1$ are characterized by the potentials
\begin{equation}\label{nu}
    \nu(i)=\sum\limits_{j=1}^{n}\alpha_j {\cal H}(f_{i,j})+\beta_j {\cal H}(-f_{i,j}).
\end{equation}

Note also that when $\nu(i)=0$, the $n$-dimensional hypercube can be readily interpreted as a system of $2d$ interacting Majorana fermions~\cite{Jia2020,Parisi1994}, or as the interaction term in the Maldacena-Qi model describing a space-time wormhole~\cite{maldacena2018}. Indeed, by introducing the gamma matrices
\begin{eqnarray}
    \gamma^L_k=\left(\bigotimes\limits_1^{k-1}\sigma_x\right)\otimes\sigma_z\otimes I_{2n-2k}, \nonumber \\
     \gamma^R_k=\left(\bigotimes\limits_1^{k-1}\sigma_x\right)\otimes\sigma_y\otimes I_{2n-2k},
\end{eqnarray}
with $\sigma_{x,y,z}$ being the Pauli matrices, the Hamiltonian in \eqref{An} can read as
\begin{eqnarray}\label{Majorana}
    H=i\sum\limits_k^d\gamma^L_k\gamma_k^R.
\end{eqnarray}
In the case when on-site potentials attain the form as in \eqref{nu}, the left and right spaces in \eqref{Majorana} become coupled. As a result, the corresponding Hamiltonian acquires a more complicated form. 

{Alternatively, the hypercube space can be mapped to the Fock space of Ising spins. Indeed, the Kronecker sum in \eqref{An} can be written as follows
\begin{eqnarray}\label{Hspin}
    H_n = \sum\limits_k^n\left(\dfrac{\alpha_k-\beta_k}{2}\sigma_z^k+\sigma_x^k\right)+I_{2n}\sum\limits_i^n\dfrac{\alpha_i+\beta_i}{2}.
\end{eqnarray}
The second term in \eqref{Hspin} just shifts the energy spectrum of the set of spins, subjected to a transverse uniform magnetic field,  described by the first sum in the Hamiltonian $H_n$.}

Finally, a general $n$-dimensional hypercube Hamiltonian in \eqref{An} can be associated with a linear network of $2^n$ coupled bosonic modes, whose frequencies are identified with potentials $\nu(i)$ [see also \eqref{Hb}].

{Evidently, hypercube graphs can be mapped onto the Hilbert space of various physical systems, including single-particle bosonic or fermionic networks, interacting Majorana fermions, and spin systems, among others.
In this work, we focus exclusively on the single-particle Hamiltonian in \eqref{Hb}.
At the same time, in what follows, we will also draw parallels with spin systems where relevant.}

\subsection{Hypercube Eigenspectrum and Symmetry}
The eigenvectors $\psi$ and eigenvalues $\lambda$ of the Hamiltonian $H_n$ in \eqref{An} are straightforwardly obtained as follows
\begin{equation}\label{eig}
\psi_{i_1,i_2,\dots,i_n}=\bigotimes\limits_{k=1}^n\psi_{i_k}^{(k)}, \quad \lambda_{i_1,i_2,\dots,i_n}=\sum\limits_{k=1}^n\lambda_{i_k}^{(k)}
\end{equation}
where $\psi^{k}_{i_k}$ ($\lambda^{k}_{i_k}$) denotes the $i_k=1,2,$ eigenvector (eigenvalue) of the matrix $S_k$ in \eqref{An}~\cite{CoxeterBook,arkhipov2023c}.
\begin{figure}[t!]
    \includegraphics[width=0.45\textwidth]{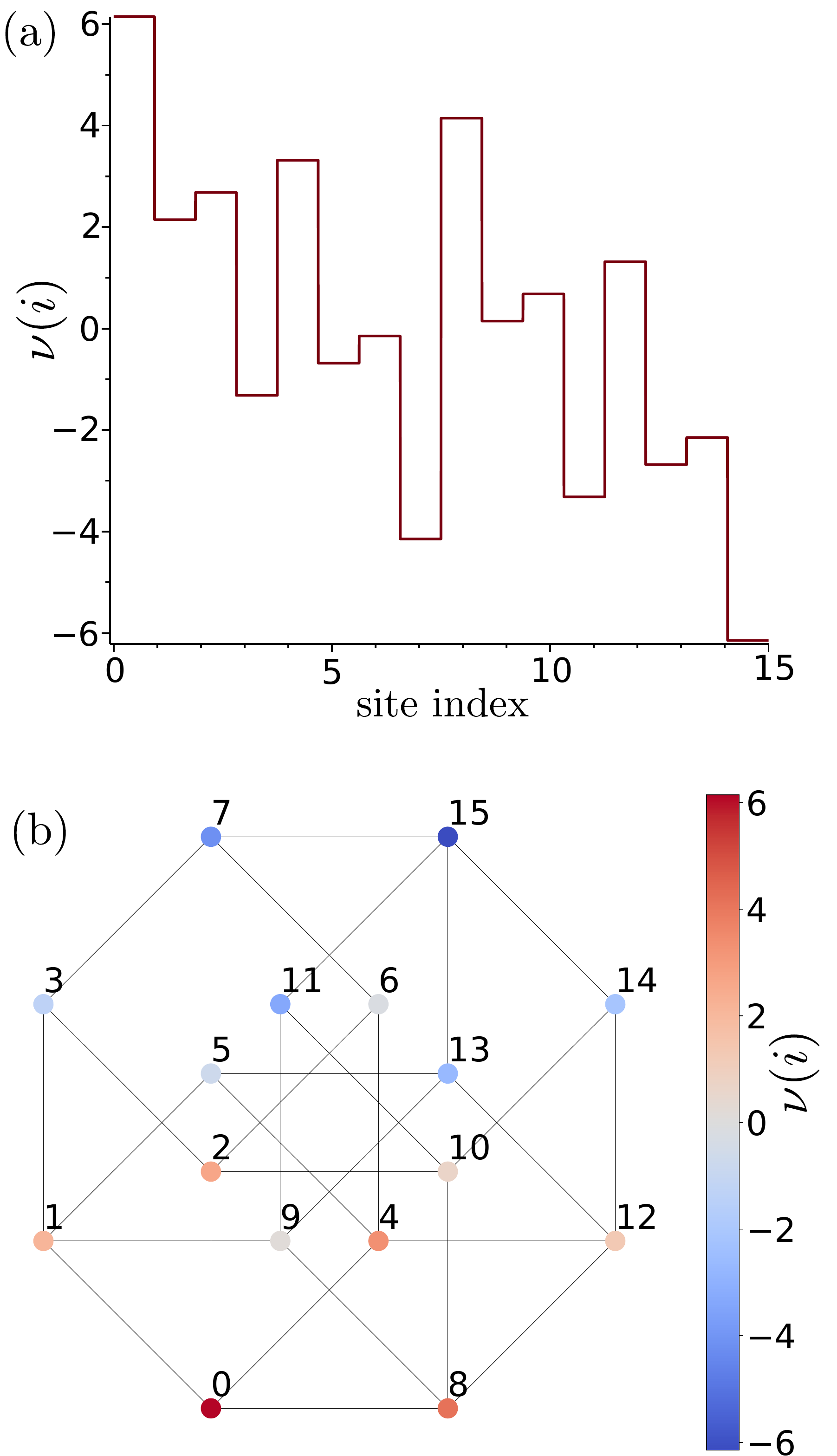}
    \caption{(a) Site potentials $\nu(i)$ on a 4D hypercube described by the Hamiltonian in \eqref{An} generated by the dion matrices $S_k$, $k=1,\dots,4$, with diagonal elements $\alpha_k=-
\beta_k=\sqrt{k}$. (b) The same site potential distribution as in panel (a) but visualized on a Petrie polygon of the given 4D hypercube. }
    \label{fig1}
\end{figure}

The eigenvectors in \eqref{eig}, thus, have a binary tree structure~\cite{Harary1988} (see also Fig.~1 in Ref.~\cite{arkhipov2023c}). Note that whereas the eigenspectrum of the $H_n$ has a tensor product structure, the underlying $2^n$ coupled bosonic modes, constituting the Hamiltonian, as in \eqref{Hb}, cannot be presented in the same manner.

The construction in \eqref{An} implies that a hypercube Hamiltonian $H_n$ possesses a chiral symmetry
(within an appropriate gauge)
\begin{equation}\label{A'}
    H'_n=H_n-\overline{\nu} I_{2^n},
\end{equation}
 where 
\begin{equation}
    \overline{\nu}=\dfrac{\max[\nu(i)]+\min[\nu(i)]}{2},
\end{equation}
with $\nu(i)$ given in \eqref{nu}, and $I$ is the identity matrix.
That is
\begin{equation}\label{C}
    {\cal C}H'_n{\cal C}^{\dagger}=-H'_n, \quad {\cal C}=\bigotimes\limits_{i=1}^n\sigma_y, \quad {\cal C}^{\dagger}{\cal C}={\cal C}{\cal C}^{\dagger}={\cal C}^{2}=I,
\end{equation}
where the symbol $\dagger$ denotes the Hermitian conjugation operation~\cite{Chiu2016}.
This chirality ensures that the eigenvalues of the matrix $H_n$ ($H'_n$) have the mirror reflection symmetry with respect to a mean eigenvalue (a zero of the energy)~\cite{arkhipov2023a}. The chiral symmetry is, in general, broken, that is ${\cal C}\psi_{\lambda_k}\equiv\psi_{\lambda_{N-1-k}}$, for ordered eigenvalues $\lambda_0<\dots<\lambda_{N-1}$. Though, non-degenerate zero-energy states of $H_n'$, if any, respect the chiral symmetry of the system~\cite{Ramachandran2017}.  
\begin{figure}[t!]
    \includegraphics[width=0.45\textwidth]{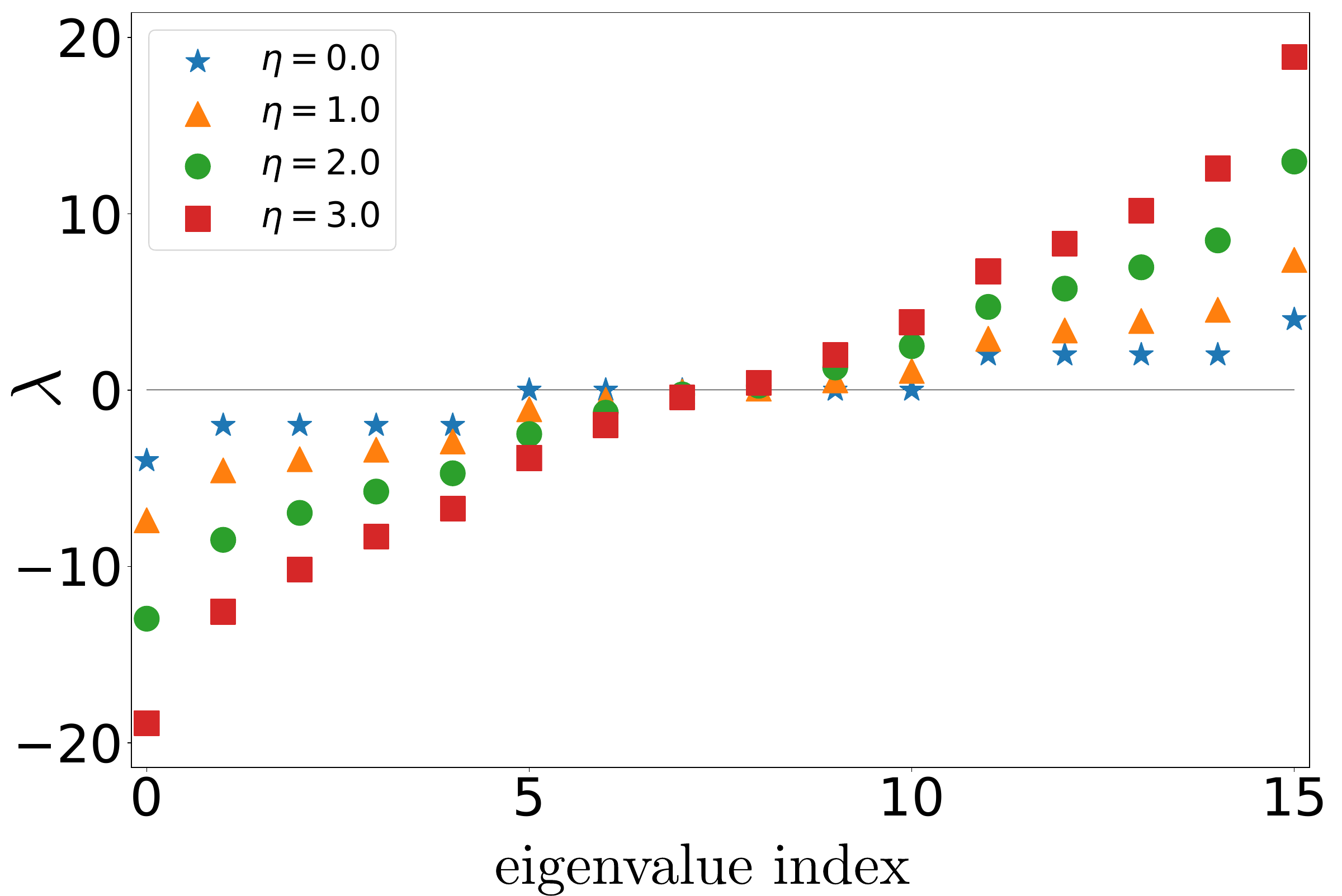}
    \caption{Eigenvalues $\lambda$  of the Hamiltonian $H_4$ describing a 4D hypercube. The matrix $H_4$ is constructed according to \eqref{An} with elements of the dion matrices $S_k$ taken as $\alpha_k=-
\beta_k=\eta\sqrt{k}$, for $k=1,2,\dots,4$, and where the constant parameter $\eta$ can acquire various values in the range $[0,1,2,3]$.}
    \label{fig2}
\end{figure}

To illustrate the described construction of a hypercube with a simple example, let us consider the $4$D case. Specifically, we assign the following values to the diagonal elements of each dion Hamiltonian matrix $S_k$ in \eqref{An}, $k=1,\dots,4$: $\beta_k=-
\alpha_k=-\sqrt{k}$, which guarantees that the on-site potentials of the resulting hypercube with $H_4$ are incommensurate.  To visualize  the obtained 4D-hypercube with given site potentials, we project it onto the 2D Petrie polygon, as shown in \figref{fig1}. A Petrie polygon for a regular hypercube or any polytope of $n$ dimensions is a skew polygon in which every $(n-1)$ consecutive sides (but not $n$) belongs to one of the facets~\cite{CoxeterBook,Nelson1984}. 
Generally, a hypercube with commensurate site potentials produces a degenerate energy spectrum, whereas incommensurate site potentials lift this degeneracy (see also \figref{fig2} and the text below).

\section{Engineering Compact localized states}\label{III}
\begin{figure}[t!]
    \includegraphics[width=0.495\textwidth]{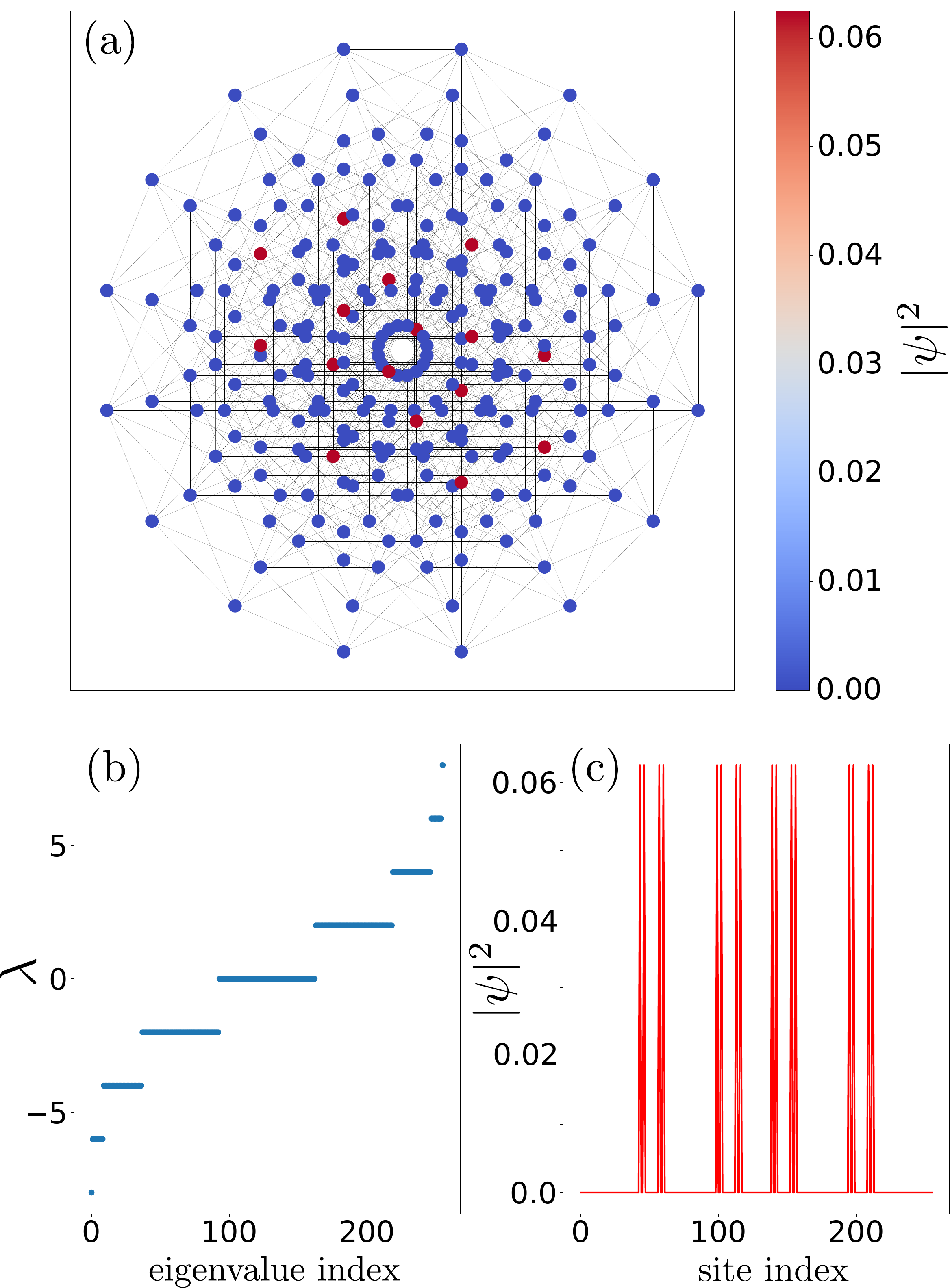}
    \caption{ (a) The intensity of a zero-energy compact localized state (CLS) of an 8D hypercube with zero site potentials visualized on a hypercube Petrie polygon. (b) The degenerate `flat band' energy spectrum of  the given hypercube Hamiltonian $H_8$. (c) Similar to panel (a) showing the CLS intensity distribution as a function of the site index. }
    \label{fig7}
\end{figure}
Elaborating on the method described above, we now show how to construct CLSs on hypercubes with sites that have constant potentials.  Without loss of generality, we assume that the constant potential is zero ($\alpha_k=\beta_k=0$  $\forall k$).
For an arbitrary $n$-dimensional hypercube, the zero potential always results in a highly degenerate spectrum, reminiscent to that observed in flat-band systems but in reciprocal space~\cite{Nori1990}. However, it is only for even ($2n$)-dimensional  hypercubes that this flat spectrum  contains  a zero-energy level, which, moreover, has a degeneracy of the degree~\cite{arkhipov2023c}
\begin{eqnarray}\label{m2n}
    m_{2n}=(2n)!/(n!)^2.
\end{eqnarray}
For instance,  in the case of the 8D hypercube with $2^8=256$ vertices and $2^7\cdot 8=1024$ edges, the degeneracy of the zero-energy level is $m_8=8!/(4!)^2=70$ [see also \figref{fig7}(b)]. 
Furthermore, these degenerate zero-energy states (ZESs), denoted as $\psi_{0}^{k}$, $k=1,\dots,m_{2n}$, are {\it extended} in nature, i.e., with site amplitudes equal $\psi_0^k(i)\equiv\pm1$~\footnote{Indeed, each matrix $S_k$ in \eqref{An} with $\alpha_k=\beta_k=0$ has two eigenstates $\psi_1=[1,1]^T$ and $\psi_2=[-1,1]^T$, corresponding to eigenvalues $\lambda_{1,2}=\pm 1$. As such, the $2n!/[n!]^2$-folded degenerate ZESs for a given Hamiltonian $H_{2n}$, are formed by $2n$-folded Kronecker products of $n!$ number of states $\psi_1$ and the same number of states $\psi_2$.}, and which form the orthogonal basis. 
Because of this, any superposition
\begin{equation}\label{super}
    \psi_0^{\rm s}\equiv\sum\limits_k {c_k}\psi_{0}^{k}, \quad c_k\in{\mathbb C},
\end{equation}
is also a ZES. Some superpositions in \eqref{super} can result in {\it destructive interference, leading thus to the formation of the CLSs}, where only a fraction of the sites have nonzero amplitudes~\cite{Nori1990}. {Indeed, since each extended state $\psi_{0}^{k}$ in \eqref{super} is formed by the multiple tensor products of two eigenvectors $(1,1)^T$ and $(-1,1)^T$ of the generating matrix $S_k=\sigma_x$, according to Eqs.~(\ref{Sk}) and (\ref{eig}), their combinations can lead to destructive interference between their entries that differ only in sign. Since one can readily construct all $(2n)!/(n!)^2$ ZESs, one can subsequently form various combinations in \eqref{super} which result in the CLSs. One can show that there at least $2^n$ such zero-energy CLSs on the $2n$-dimensional hypercube. We elaborate on this in more detail in Appendix~\ref{AP1}.}

We plot one of the realizations of such CLSs in \figref{fig7}. As can be seen from the graph, there are only a few nonzero peaks in the CLS, whereas other sites have strictly zero amplitude. Clearly, one can also find such a similarity transformation (a permutation matrix) for the Hamiltonian $H_8$  which can reshuffle these nonzero peaks of the ZES~\cite{Rontgen2018}.

Note that the Hamiltonian of an $n$-dimensional bosonic hypercube graph in \eqref{Hb} with everywhere $\nu(i)=0$, apart from the chiral symmetry, also respects parity symmetry, meaning that $H_n={\cal P}H_n{\cal P}$, where the parity operator ${\cal P}=\bigotimes\limits_{k=1}^n\sigma_x$. Moreover, the zero-energy degenerate states $\psi_0^k$ share the same parity, implying that the superpositions in \eqref{super} are also eigenstates of the operator $\cal P$. 

Regarding chiral symmetry, the degenerate states $\psi_0^k$ may enter the combination in \eqref{super} with opposite chirality; meaning that, in general, a superposition $\psi_0^{\rm s}$ is not the eigenstate of the operator $\cal C$. However, certain combinations of $\psi_0^k$ may form chiral eigenstates $\psi_0^{\rm s}$: ${\cal C}\psi_0^{\rm s}\equiv\psi_0^{\rm s}$, making them additionally robust against perturbations that respect this symmetry~\cite{Ramachandran2017}. 

\subsection{Effects of perturbations on stability of compact localized states}
\subsubsection{Correlated perturbations}
{The appearance of the high degeneracy in the hypercube spectrum, discussed above, can also be easily understood when mapping it to the Fock space of a system of {\it identical} spins subjected to a uniform magnetic field, according to \eqref{Hspin}. Moreover, by introducing hypercube perturbations in the form 
\begin{eqnarray}\label{Hspin1}
    H_{\rm pert}=\sum\sigma_x^{i}\sigma_x^{i+1}\dots\sigma_x^{i+2k}, 
\end{eqnarray}
mimicking odd-spin interactions ($k\in\mathbb{Z}^+$) in the context of interacting spin systems,  one can preserve the CLSs set from the unperturbed case, since such perturbation commutes with the unperturbed Hamiltonian $H=\sum \sigma_x$. In a particular case when $H_{\rm pert}={\cal P}$, these perturbations introduce extra edges between hypercube vertices, forming the hypercube diagonals. Remarkably, despite significantly altering the link topology, these modifications leave the hypercube eigenfunction space unchanged. In the single-particle framework, such diagonal links correspond to additional intermode couplings in the bosonic networks, according to \eqref{Hb}.
}

{From the above analysis it follows that  perturbations respecting both chiral and parity symmetries of the hypercube Hamiltonian can preserve the presence of zero-energy CLSs, formed by the eigenstates of $\cal P$ and $\cal C$ operators. This observation strongly resonates with the findings in Refs.~\cite{Karle2021,Turner2018,Schecter2018}, where an exponentially large nullspace, featuring localization of the many-body interacting spin system, is attributed to the presence of parity and chiral symmetries in the system's Hamiltonian. 
Specifically, for a perturbed hypercube Hamiltonian in the form $H\to H+H_{\rm pert}$, where 
\begin{eqnarray}\label{Hspin2}
    H_{\rm pert}=&&a\sum (\sigma_y^{i}\sigma_z^{i+1}+\sigma_z^{i}\sigma_y^{i+1})+b\sum \sigma_z^{i-1}\sigma_x^{i}\sigma_z^{i+1} \nonumber \\ 
    &&+c\sum\sigma_y^{i-1}\sigma_x^{i}\sigma_y^{i+1},
\end{eqnarray}
the hypercube spectrum continues to host degenerate ZESs, and hence CLSs, since the perturbation in \eqref{Hspin2} satisfies $[H_{\rm pert},{\cal P}]=0$, and $\{H_{\rm pert},{\cal C}\}=0$.} 

{The `two-spin interaction' term in \eqref{Hspin2} modifies only the non-zero edge weights of the unperturbed hypercube by adding imaginary values ($\pm ia$), leaving the link topology unchanged. 
 The second sum also alters the weights (with real values $\pm b$) of the unperturbed hypercube edges, but can reduce the degree of certain vertices by one if a modified link weight becomes zero. The degree of a vertex of a graph is the number of edges that are incident to the vertex. For instance, for an $n$-dimensional hypercube the degree of each vertex is $n$.} 
 
 {In contrast, the third sum always increases the degree of certain vertices by one by assigning the weight ($\pm c$) to newly formed edges; thus, leading to a modification of the hypercube link network. This modification, again, is simply reflected in the change of the corresponding mode couplings in the single-particle Hamiltonian in \eqref{Hb}.
}

{We note that the method described for realizing CLSs also echoes the approach studied in Ref.~\cite{Danieli2020}, which, in particular, analyzes bosonic and spinful fermionic many-body flat-band Hamiltonians. This similarity arises when mapping the Fock space of such Hamiltonians onto the hypercube space, where the local and global integrals of motion of the many-body system are expressed as local and global parity symmetries on the associated hypercube graph. Together with the chiral symmetry of the hypercube, this leads to the existence of zero-energy CLSs, as revealed above. Thus, the hypercube framework allows to unify different approaches~\cite{Karle2021,Danieli2020} developed for constructing CLSs.
}

\subsubsection{Effects of pure disorder on CLSs}
The effects of uncorrelated disorder on hypercubes, with initial zero-energy site potentials, though {\it implicitly}, have been studied in Ref.~\cite{Laumann2014} in the context of quantum random energy models. That is, a perturbed hypercube Hamiltonian in \eqref{Hspin} can be written as~\cite{Laumann2014}
\begin{eqnarray}\label{Hdis}
    H = E(\{\sigma_z^i\})+\sum\limits\sigma_x^i,
\end{eqnarray}
where the first term represents a random operator assigning hypercube vertex potentials with values drawn from a Gaussian distribution.

{Apparently, any non-zero disorder destroys CLSs by breaking parity and chiral symmetries of the hypercube, although the eigenstates remain delocalized for sufficiently small perturbations. As disorder is increased further, the hypercube graph undergoes a delocalization-localization transition. 
Denoting the energy density of hypercube states as $\epsilon$, the transition for the Hamiltonian in \eqref{Hdis} occurs at $\epsilon = 1$~\cite{Laumann2014}, with the condition $\epsilon > 1$ marking the onset of localization. In the context of spin systems, this corresponds to the hypercube network entering a many-body localized phase. This disorder-induced localized regime is characterized solely by the presence of NCLSs.
}

\section{Engineering non-compact localized states}\label{IV}
Here we discuss the formation of zero-energy NCLSs on hypercubes with both commensurate and incommensurate site potentials. Additionally, we demonstrate how to generate such states with specific localization features.
%FIGURE 3
\begin{figure*}[t!]
   \includegraphics[width=0.99\textwidth]{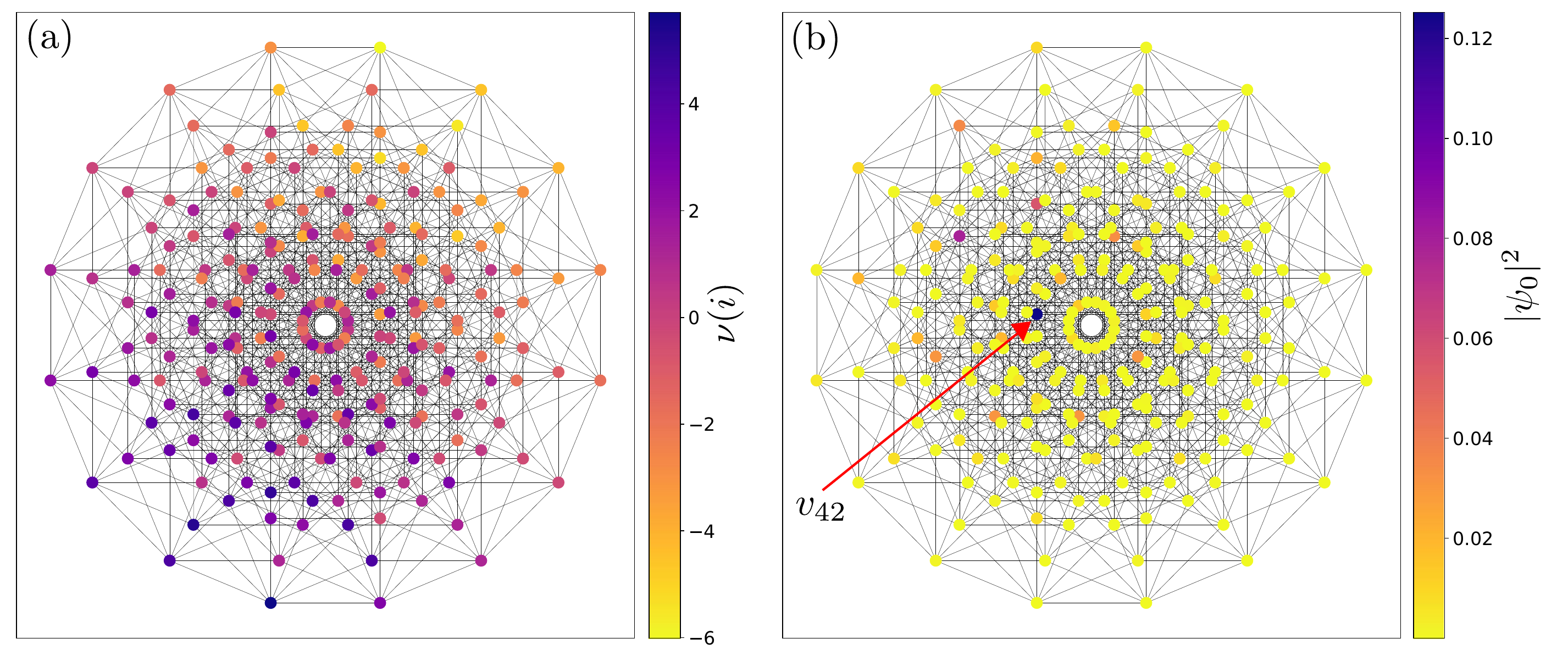}
    \includegraphics[width=0.99\textwidth]{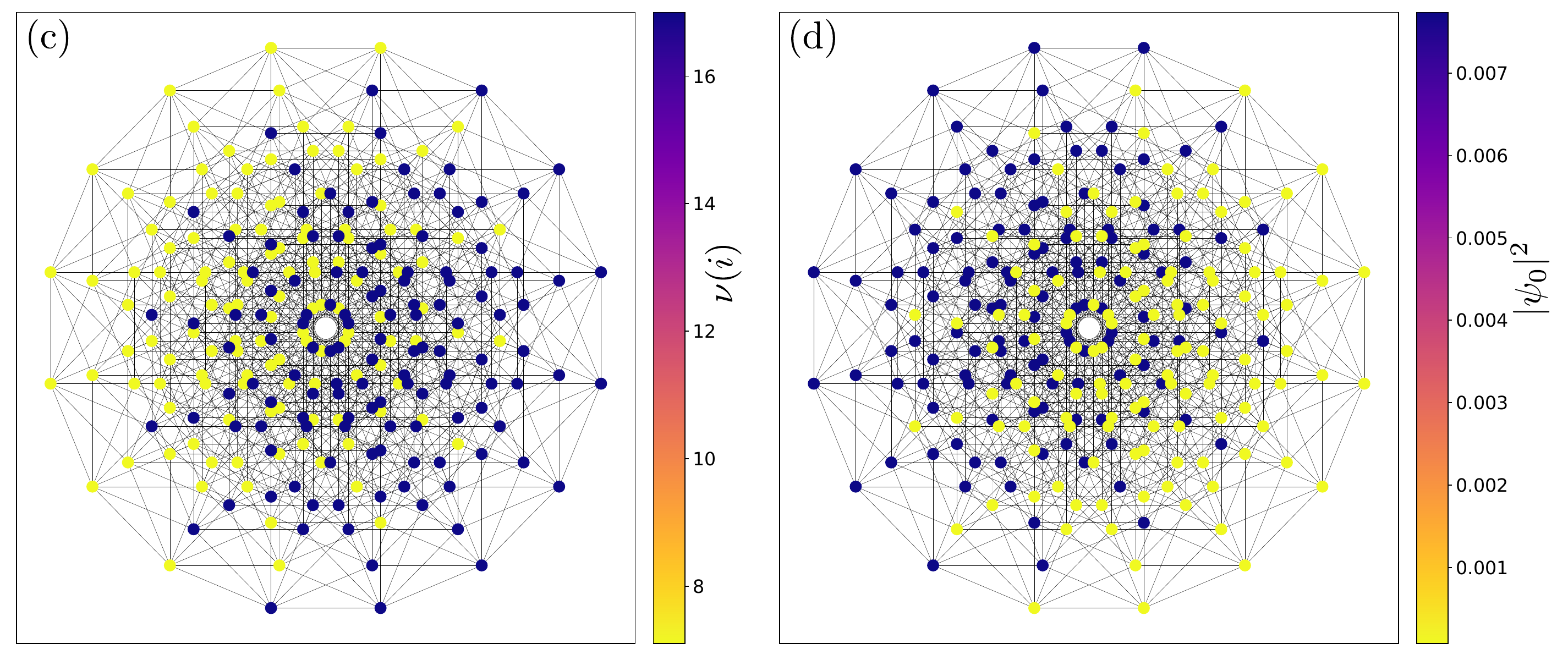}
    \caption{Panels (a) and (c) show the site potential distributions of the 8D hypercube, projected in 2D, that result in (b) the zero-energy non-compact localized state (NCLS) $\psi^{42}$, and (d) the domain wall NCLS, respectively.}
    \label{fig3}
\end{figure*}

A ZES of a Hamiltonian $H_n$ describing a $n$-dimensional hypercube, with varying site potentials in \eqref{nu}, can always be straightforwardly obtained from \eqref{An} by choosing the parameters of each matrix $S_k$ such that $\beta_k=\alpha_k^{-1}$. This is because $S_k$ has a zero-valued determinant whenever $\alpha_k\beta_k=1$.
The (unnormalized) zero-energy eigenvectors of the $S_k$ reads~\footnote{The other eigenvector with the eigenvalue $\lambda_k=\alpha_k+\alpha_k^{-1}$ attains the form $\psi_{k,\lambda\neq0}\equiv\Big[\alpha_k, 1\Big]^T$.}
\begin{equation}\label{psik0}
    \psi_{k,\lambda=0}\equiv\Big[\beta_k,1\Big]^T.
\end{equation}
For a $n$-dimensional hypercube, the zero-energy eigenvector, according to \eqref{eig}, then takes the form
\begin{equation}\label{psi0}
    \psi_{0}=\bigotimes\limits_{k=1}^n\psi_{k,\lambda=0}.
\end{equation}
Due to the binary-tree structure of the eigenvectors, resulting from the Kronecker product~\cite{arkhipov2023c}, the elements of the resulting eigenstate $\psi_0$ can be easily encoded with binary strings. Specifically, the binary string consisting of $n$ elements $i=i_1i_2\dots i_n$, with $i_k=0,1$, represents the decimal index $i$ of the site amplitude $\psi_0(i)$.  Moreover, this string also encrypts the element $\psi_0(i)$ expressed as 
\begin{equation}\label{bool}
    \psi_0(i)\equiv\beta_1^{\neg i_1}\beta_2^{\neg i_2}\dots \beta_D^{\neg i_n},
\end{equation} 
where the notation $\neg i_k$ denotes the NOT operation over the boolean $i_k$.
For instance, take a $3$D hypercube or simply a cube. Its zero-energy eigenvector reads
\begin{eqnarray}\label{psi3}
    \psi_0^{(\rm b)}\equiv\left[\stackon{000}{\beta_1\beta_2\beta_3,},\stackon{001}{\beta_1\beta_2,},\stackon{010}{\beta_1\beta_3,},\stackon{011}{\beta_1,},\stackon{100}{\beta_2\beta_3,},\stackon{101}{\beta_2,},\stackon{110}{\beta_3,}, \stackon{111}{1} \right]^T. \nonumber \\ 
\end{eqnarray}
In \eqref{psi3}, the upper line accounts for the actual elements of the vector $\psi_0$, whereas the lower line refers to their binary representation according to \eqref{bool}.

\begin{figure}
    \includegraphics[width=0.49\textwidth]{fig5.pdf}
   \caption{{Energy spectrum of the perturbed hypercube Hamiltonian and its zero-energy state intensity for various perturbations drawn from a uniform distribution in the range $[-W,W]$ (see details in the main text). 
  Energy spectrum of the Hamiltonian which generates (a) a zero-energy localized state $\psi^{42}$ and (c) domain wall state, whose state intensities are shown in panels (b) and (d), respectively.}  The inset in panel (a) shows the detailed energy spectrum [which is almost overlapped for different values of the parameter $W$] and corresponding perturbed zero-energy state around the zero-energy level.}   \label{fig5}
\end{figure}

Based on the structure of the {\it zero-energy eigenvector} in \eqref{psi0}, one can construct a vector state of an $n$-dimensional hypercube with {\it desired} localization characteristics. 
Indeed,  suppose that one wishes to obtain a specific zero-energy eigenstate on an 8D hypercube.  
Assume first that the desired ZES is a single-site localized state at the vertex, say $v(42)$. In the state $\psi^{42}_0$, the vertex with the index 
 $42$, which in the boolean form reads as ${\rm b}00101010$, has the following amplitude 
$\psi_0^{42} = \beta_1\beta_2\beta_4\beta_6\beta_8$, according to \eqref{bool}.
To guarantee that the ZES amplitude concentrates at the vertex $v(42)$ one must then ensure that $|\beta_{3,5,7}|\ll1< |\beta_{1,2,4,6,8}|$. We plot one of the possible  hypercube potential distributions on \figref{fig3}(a) which generates the  localized state shown in \figref{fig3}(b).  We also present complementary plots with the eigenspectrum decomposition for the $8$D cube in \figref{fig5}. From \figref{fig5} it is seen that apart from a delta-like peak at the vertex $v(42)$, there are a number of other smaller peaks which correspond to the products of four, three, and so on, terms of $\beta_k$ in \eqref{bool}. By varying the values of $\beta_k$ one can correspondingly modify the intensity of these satellite peaks as desired.
It must be noted that the localization is observed in the whole eigenspectrum, not only for the ZES.

Next, suppose that the intended ZES is a ``domain wall'' state, meaning that only one half (assume the first half) of the vertices can be excited~\cite{Choi2016}. This can be readily achieved by having the first half of the elements of the zero-energy vector $\psi_0$ possess the term $\beta_1$. Note that each term $\beta_k$ has a vertex periodicity $2^{-k}L$ in $\psi_0$, where $L=2^n$ is the system length, in the considered case $L=2^8$. By setting $|\beta_1|>1$, and the remaining terms $|\beta_{2,\dots,8}|\approx 1$, one attains an 8D-hypercube with the desired ZES [see Figs.~\ref{fig3}(c),(d)]. The explicit form of the spectrum and the density of the domain wall state $\psi_0$  are also shown in Figs.~\ref{fig5}(c),(d). Panels (c) and (d) in \figref{fig3} show that vertex excitations in this ZES exhibit a duality with the potential distribution on the hypercube. Specifically, hypercube sites with smaller potentials (in absolute magnitude) have a higher probability of being excited in this state.

\subsection{Robustness of the non-compact localized states to disorder}\label{IVA}
{Here, we analyze the effects of both uncorrelated and correlated disorder on the stability of the engineered NCLSs discussed above.}
\subsubsection{Uncorrelated disorder}\label{V1}

{We start our consideration from the effects of pure, i.e., uncorrelated, disorder on the stability of the NCLSs.
Such random perturbations are imposed on all diagonal elements of the hypercube matrix, similar to \eqref{Hdis}, as used in the study of quantum random energy models~\cite{Laumann2014,Baldwin2016}.}

We study the {\it robustness} of the 8D hypercube Hamiltonian when the site potentials are perturbed by random values drawn from a uniform distribution over the range $[-W,W]$, with $W\in{\mathbb R}$.  
Two scenarios can be distinguished upon such perturbations: (i) when the density of states near zero energy is high, and (ii) when the ZES is well isolated from the rest of the spectrum. 

For the first case, our analysis shows that whenever  $W\lesssim|\lambda^{\rm e}|$, 
where $\lambda^{\rm e}$ is the energy gap, i.e., the distance between the first excited state above or below zero of the unperturbed system, the initial ZES remains immune to perturbations. Namely, despite the fact that the zero energy of the initial state can be shifted, the perturbed ZES remains closest to the zero energy level [see \figref{fig5}(a),(b)]. However, for larger values of $W$, the initial ZES, while preserving its shape, can be shifted far away from the level $\lambda=0$, becoming an `excited' state [see \figref{fig5}(a),(b)].

For the second scenario, the ZES exhibits larger robustness against disorder. Namely, the energy of the initial ZES remains zero upon perturbations. However, the modification of the state increases with larger disorder $W$ [see panels (c) and (d) in \figref{fig5}]. In both scenarios, the increasing values of $W$ eventually lead to the emergence of continuous energy spectra, which is an indicator of a completely disordered system~\cite{Pastur1980} [see \figref{fig5}]. 

The above observation implies that  a ZES of a hypercube,  within a region of the state space with a high (low) density of states,  exhibits high (low) susceptibility to perturbations. In other words, isolating the ZES in the system spectrum allows for the engineering of the robust system response in the presence of disorder. 
We also note that the ZES remains immune to purely imaginary perturbations, e.g.,  when the potentials disorder is dissipative in nature (see Appendix~\ref{AA} for details).

\subsubsection{Correlated disorder}
{The previous analysis suggests that any correlated disorder represented, e.g., by a perturbing hypercube Hamiltonian in the form $H\to H+ H_{\rm pert}$, where
\begin{eqnarray}\label{Hncls}
    H_{\rm pert}\equiv\sum\limits_{i,k}\epsilon_i\sigma_z^{i}\sigma_z^{i+1}\dots\sigma_z^{i+k},
\end{eqnarray}
 also preserves the structure of the initial NCLSs. Such perturbations modify the hypercube vertex potentials only, similar to the case discussed earlier on uncorrelated disorder. The random perturbation parameters $\epsilon_i$ in \eqref{Hncls} can be taken from a uniform distribution $[-W,W]$, similar to that in \figref{fig5}.} 

{The fact that the vertex potential disorder, when small enough, does not significantly affect the hypercube NCLSs can be also understood in the context of interacting spin systems, as directly indicated by \eqref{Hncls}.  Indeed, when mapping the hypercube network onto the Fock space of the spin system, the spin interactions in \eqref{Hncls} do not induce excitation transfer between lattice sites~\cite{Danieli2024}. Consequently, such perturbations tend to preserve the shape of the initial hypercube NCLSs. }

\section{Discussion}\label{V}
{In analyzing the robustness of NCLSs against disorder, it is useful to draw parallels with previous studies on {\it disorder}-induced Fock-space many-body localization in spin systems~\cite{Laumann2014,Roy2020mbl}. Notably, as indicated in Ref.~\cite{Roy2020mbl}, disorder-induced many-body localized states (DIMBLS) remain stable when Fock-space site energies exhibit maximal correlations at finite Hamming distances. We recall that the Hamming distance between two nodes on the graph is defined as the shortest path between them following the links. This condition is exactly satisfied in our case. It can be said that the NCLSs studied here represent an extreme case of the DIMBLS, since one sets specific (not necessarily random) values of $\beta_i$ in \eqref{bool} to construct a given localized state. The construction in \eqref{bool} maximizes correlations between hypercube sites, thus ensuring the NCLSs robustness.} 

{The NCLSs stability to pure disorder analyzed in Sec.~\ref{V1} can be also understood in this context. Indeed, the pure disorder introduced  on the hypercube vertices, similar to \eqref{Hdis}, competes with the inherent hypercube site correlations determined by the NCLS construction in \eqref{bool}. However, the NCLS are already localized, so introducing uncorrelated disorder merely modifies their shape and preserves the localization phase on the hypercube graph~\cite{Roy2020mbl}.}

\section{Conclusions and Outlook}\label{VI}
{In conclusion, we reveal that localization phenomena can naturally emerge on hypercube  graphs without disorder.
At the same time, we presented a method allowing for engineering robust CLSs and NCLSs on such disorder-free hypercubes.} 
Namely, we showed that for the hypercubes with constant site potentials, the resulting highly degenerate energy spectrum in real space enables producing CLSs. Whereas the   incommensurate site potentials lead to the emergence of NCLSs with prescribed localization features.
Given the importance of CLSs and NCLSs in the realization of various information and wave manipulation protocols, our results can potentially lead to advancements in these fields. The hypercube graphs presented and their localization properties can be directly simulated in existing experimental photonic platforms exploiting both real and synthetic spaces, e.g., networks of coupled cavities or waveguides~\cite{Leefmans2022,Parto2023,Leefmans2024}.

{Moreover, our findings indicate that hypercube structures can be directly associated with the Fock space of  interacting spin systems. This provides additional insights into many-body localized states and can open new avenues for simulating complex quantum many-body models. That is, the demonstrated robustness of constructed CLSs and NCLSs against hypercube perturbations, which effectively emulate many-body interaction in the Fock space of spin systems, offers a new perspective on the origin of many-body localization in such systems  }

Additionally, the approach used can be readily extended to other hyperpolytopes constructed by iterative Cartan products of triangles, tetrahedrons, and so on. Consequently, in future research we wish to explore other types of CLSs and NCLSs that can be engineered in these hyperstructures. 
In relation to this, it would be also interesting to investigate how the studied localization phenomena on hypercubes are modified when mapped to lower-dimensional systems~\cite{arkhipov2023a}.

{Analogous to the hypercubes studied here, the hyperpolytopes could be also potentially mapped to the Fock space of higher-spin interacting models. This, in turn, can shed light not only on localization phenomena in more complex quantum many-body systems but also can lead to their effective simulations with linear hyperpolytope networks.}

\acknowledgements
I.A. acknowledges support from Air Force Office of Scientific Research (AFOSR) Award No. FA8655-24-1-7376, and 
from the Ministry of Education, Youth and Sports of the Czech Republic Grant OP JAC No. CZ.02.01.01/00/22\_008/0004596.
F.N. is supported in part by:
the Japan Science and Technology Agency (JST)
[via the CREST Quantum Frontiers program Grant No. JPMJCR24I2,
the Quantum Leap Flagship Program (Q-LEAP), and the Moonshot R\&D Grant Number JPMJMS2061], and the Office of Naval Research (ONR) Global (via Grant No. N62909-23-1-2074).

\appendix

\section{Another method to generate  zero-energy compact localized states in Sec.~\ref{III}}\label{AP1}
In Sec.~\ref{III} we mentioned that zero-energy CLSs can be simply generated by various combinations of the ZES $\psi_0^k$ in \eqref{super}. These ZES $\psi_0^k$ of the $2n$-dimensional hypercube are obtained via tensor products of two (unnormalized) eigenvectors  $\phi_1$ and $\phi_2$ of the generating matrix $S=\sigma_x$. Namely,
\begin{eqnarray}\label{psi0k}
     \psi_{0}^k=\bigotimes\limits_{i, j_i=\{1,2\}}^{2n}\phi_{j_i}, \quad k=1,\dots,2n!/(n!)^2
\end{eqnarray}
where
\begin{eqnarray}
    \phi_1\equiv[1,1]^T, \quad \phi_2\equiv[-1,1]^T.
\end{eqnarray}
The tensor product in \eqref{psi0k} contains $n$ number of vectors $\phi_1$ and the same number of vectors $\phi_2$ in various combinations, ensuring that the resulting vector $\psi_0^k$ is a ZES, according to \eqref{eig}. 

As a result, the elements of the (unnormalized) ZESs $\psi_0^k$ consist of an equal number of $\pm1$ entries. The precise position of $\pm1$ values in a vector $\psi_0^k$ is determined by the expression in \eqref{bool}, where each $\beta_{i}=\pm1$ is defined by the first element of the $i$th eigenvector $\phi_{j_i=1,2}$ in the product in \eqref{psi0k}.
Clearly, because of the latter, one can realize various CLSs through symmetric and asymmetric combinations of the eigenstates $\psi_0^k$, according to \eqref{super}. However, while this approach can, in principle, yield CLSs, it would most likely require numerical methods, especially for high-dimensional hypercubes, since the specific combinations that form CLSs are generally unknown.

Alternatively, CLSs can be generated using a different approach. By noting that the ZESs of a 
$2n$-dimensional hypercube can also arise from the tensor products of ZESs of lower-dimensional hypercubes ($<2n$), one can first construct CLSs for these lower-dimensional hypercubes and then obtain CLSs for the 
$2n$-dimensional case through their corresponding tensor products.

Let us demonstrate the latter on a simple example of a $4$D hypercube.
The sixteen eigenstates of the $4$D hypercube are obtained from \eqref{psi0k}, through four-fold tensor products of the two states $\phi_{1,2}$. However, the lower-dimensional hypercube with respect to the 4D hypercube, which still possesses ZESs, is a 2-dimensional hypercube, i.e., a square. The square is characterized by two ZESs, namely, $\psi_0^1=\phi_1\otimes\phi_2$ and $\psi_0^2=\phi_2\otimes\phi_1$. More explicitly,
\begin{eqnarray}\label{sq_eig}
    \psi_0^1=\begin{pmatrix}
        -1 \\
        1 \\
        -1 \\
        1
    \end{pmatrix}, \quad \psi_0^2=\begin{pmatrix}
        -1 \\
        -1 \\
        1 \\
        1
    \end{pmatrix}.
\end{eqnarray}
One can form now CLS-like zero-energy states through symmetric and asymmetric combinations of the vectors in \eqref{sq_eig}. Namely,
\begin{eqnarray}\label{sq_eig}
    \psi_1^{\rm s}\equiv\psi_0^1+\psi_0^2\equiv\begin{pmatrix}
        -1 \\
        0 \\
        0 \\
        1
    \end{pmatrix}, \quad  \psi_2^{\rm s}\equiv\psi_0^1-\psi_0^2\equiv\begin{pmatrix}
        0 \\
        1 \\
        -1 \\
        0
    \end{pmatrix}. \nonumber \\
\end{eqnarray}
Evidently, various tensor products of these two ZESs $\psi_{1,2}^{\rm s}$ (four in total), will produce four degenerate ZESs of the $4$D hypercube, according to \eqref{m2n}. However, since the ZESs $\psi_{1,2}^{\rm s}$ already contain zero-valued elements, their tensor products further reduce the number of non-zero entries. This, in turn, facilitates the formation of CLSs in the degenerate eigenspace of higher-dimensional hypercubes. Specifically, for the 4D hypercube, one of the ZESs, obtained from the tensor product 
\begin{eqnarray}
    \psi_{1}^{\rm s}\otimes\psi_{2}^{\rm s}=[0,-1,1,0,0,0,0,0,0,0,0,0,0,1,-1,0]^T, \nonumber \\
\end{eqnarray}
form a CLS. 

Clearly, the CLSs eigenspace for any $2n$-dimensional hypercube can be constructed in this manner. Consequently, for a given $2n$-dimensional hypercube, with zero (constant) on-site potentials, there are {\it at least} $N_{\rm CLS}=2^n$ zero-energy CLSs, formed by the $n$-folded tensor products of the two eigenvectors $\psi_{1}^{\rm s}$ and $\psi_{2}^{\rm s}$ in \eqref{sq_eig}.

\section{Realizing disorder-induced non-compact localized states on disorder-free hypercubes}\label{AB}
\begin{figure}[t!]
    \includegraphics[width=0.495\textwidth]{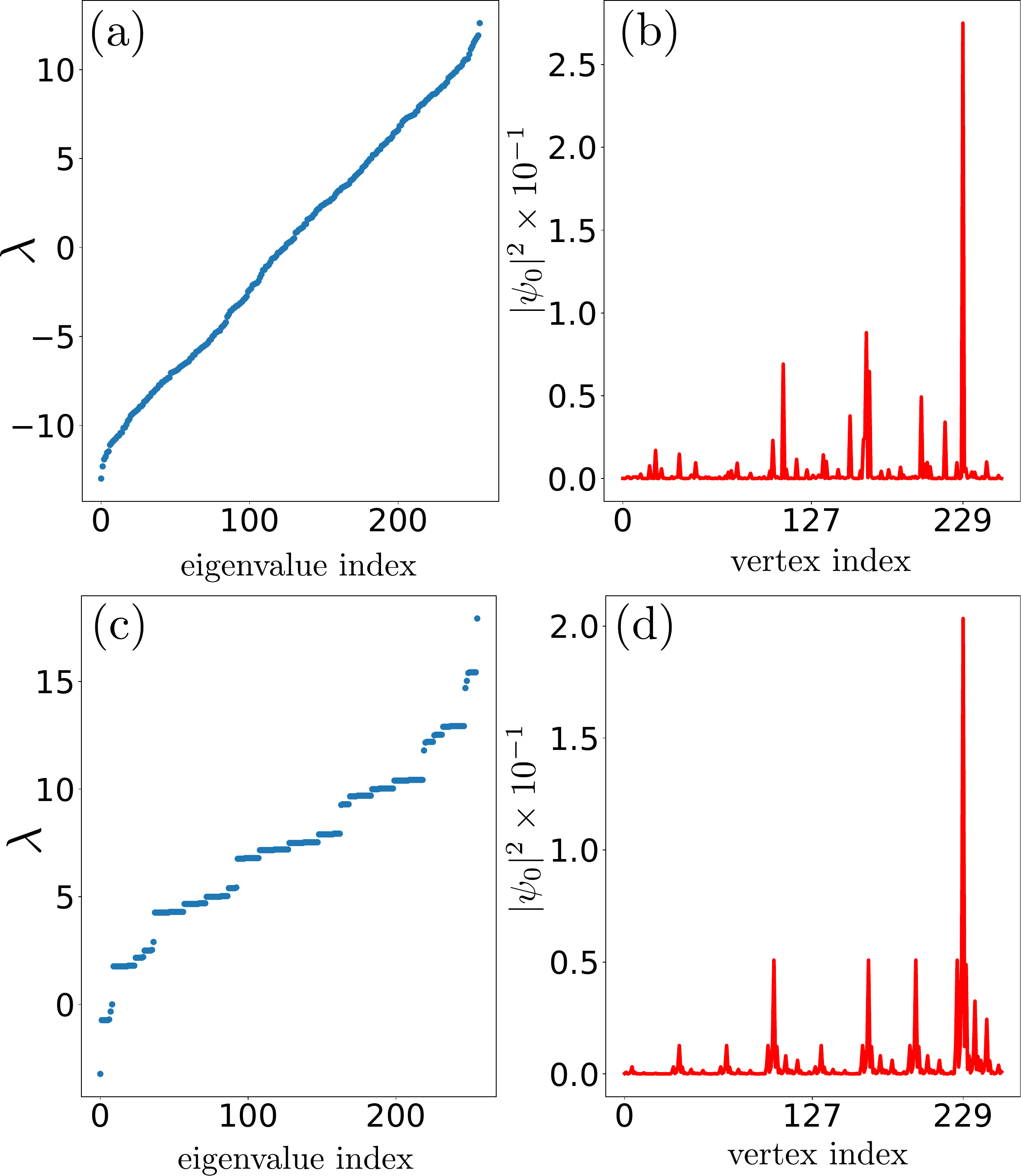}
    \caption{Energy spectrum (a) and  intensity of the localized state (b) that is closest to zero-energy of an 8D hypercube with disorder strength $W=10$. The energy spectrum (c) and zero-energy localized state intensity (d) of disorder-free 8D hypercube, constructed according to the method described in Sec.~\ref{IV} in the main text.}
    \label{S1}
\end{figure}
In the main text, we discussed the formation of NCLSs on disorder-free hypercubes. {Here we further elaborate on the localization on disorder-free hypercube graphs.} 

For that we first impose disorder on an 8D hypercube, initially characterized by zero on-site potentials, which is drawn from the uniform distribution $[-W,W]$, similar to that in Sec.~\ref{IVA} in the main text. {For large values $W\gg0$, the disorder induces localization.} We plot a disorder-induced localized state closest state to the zero-energy in \figref{S1}.  {As seen in panel (a), the energy spectrum is continuous, indicating the onset of the localization transition.} In panel (b) we show the closest to zero-energy state (ZES) that has localization peaks over a few sites, and with the maximum at at the site $v(229)$. 

Next, we aim to simulate a similar localized state using a disorder-free hypercube according to the method described in Sec.~\ref{IV}. Namely, by identifying the ZES amplitude at the site $v(229)$ as $\psi_0(229)=\beta_4\beta_5\beta_7$, according to \eqref{bool}, and by setting $|\beta_{4,5,7}|\gg1$ and $|\beta_{1,2,3,6,8}|\ll 1$, one can easily construct the ZES with similar localization characteristics [compare panels (b) and (d) in \figref{S1}], and whose spectrum is still discrete. 

\section{Robustness of non-compact localized states under dissipative perturbations}\label{AA}

In this section we discuss the robustness of NCLSs against perturbations which are dissipative in nature.
We analyze this resilience explicitly in the time evolution of the state.

\begin{figure}[t!]
    \includegraphics[width=0.48\textwidth]{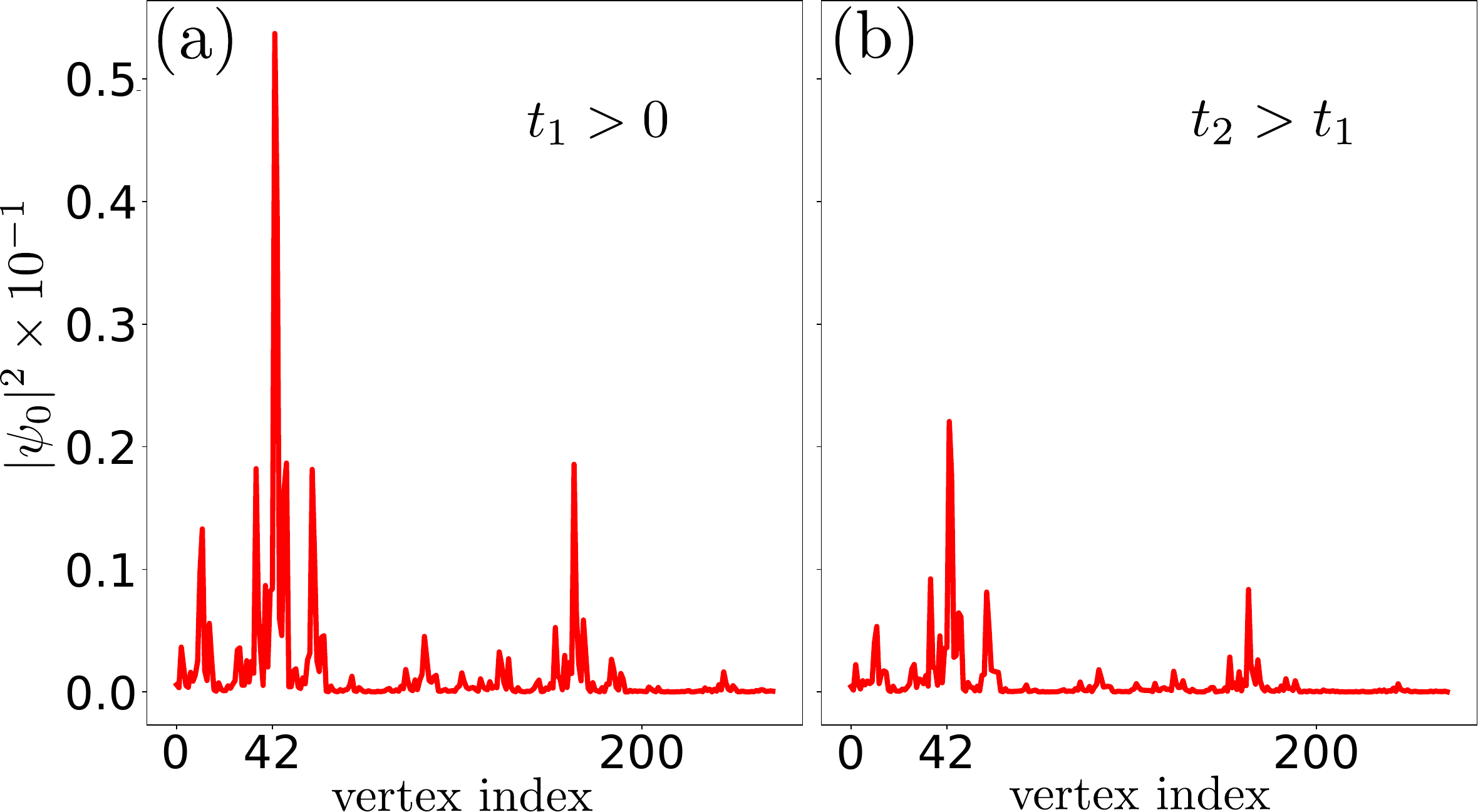}
    \caption{Time evolution of the NCLS $\psi^{42}$, initially defined on disorder-free hypercube in \figref{fig5}(b), subjected to imaginary disorder drawn from the uniform distribution $[0,i$] at time: (a) $t=1$, and (b) $t=2$. The energy and time scale is set by $\kappa=1$ in \eqref{An}}
    \label{S2}
\end{figure}

Indeed, let us now assume that the state dynamics is governed by the time-independent Schr\"odinger equation, with the (perturbed) Hamiltonian $H_n$. The hypercube Hamiltonian can describe a  network of coupled optical cavities or waveguides with losses, and thus can be written in the matrix mode representation.  The solution for the state dynamics then reads 
\begin{equation}
    \psi(t)=\exp\Big(-iH_nt\Big)\psi(0),
\end{equation}
where we set the Planck's constant $\hbar=1$.
 For the studied 8D hypercube, the Hamiltonian matrix takes the form
 \begin{equation}
     H_n=H_8-i{\rm diag}[\gamma_{1},\dots,\gamma_{N}],
 \end{equation}
 where $H_8$ is the Hamiltonian of the unperturbed  hypercube in \eqref{An}, and  the parameters $\gamma_{k}$ are sampled from the uniform distribution $[0,W]$.  We initialize the system in the zero-energy state $\psi^{42}$, localized at the vertex $v(42)$ as shown in Figs.~\ref{fig3}(b), \ref{fig5}(b), and we set $W=1$. In this case, the state $\psi^{42}$ still preserves its shape over time, though it begins steadily decaying with a certain rate depending on the disorder strength (see \figref{S2}). For larger values $W\gg1$, the dynamics of the state becomes more intricate due to the growing role of the dissipation eventually leading to the complete loss of the initial state.

 \bibliography{references}

%merlin.mbs apsrev4-1.bst 2010-07-25 4.21a (PWD, AO, DPC) hacked
%Control: key (0)
%Control: author (0) dotless jnrlst
%Control: editor formatted (1) identically to author
%Control: production of article title (0) allowed
%Control: page (1) range
%Control: year (0) verbatim
%Control: production of eprint (0) enabled
\begin{thebibliography}{73}%
\makeatletter
\providecommand \@ifxundefined [1]{%
 \@ifx{#1\undefined}
}%
\providecommand \@ifnum [1]{%
 \ifnum #1\expandafter \@firstoftwo
 \else \expandafter \@secondoftwo
 \fi
}%
\providecommand \@ifx [1]{%
 \ifx #1\expandafter \@firstoftwo
 \else \expandafter \@secondoftwo
 \fi
}%
\providecommand \natexlab [1]{#1}%
\providecommand \enquote  [1]{``#1''}%
\providecommand \bibnamefont  [1]{#1}%
\providecommand \bibfnamefont [1]{#1}%
\providecommand \citenamefont [1]{#1}%
\providecommand \href@noop [0]{\@secondoftwo}%
\providecommand \href [0]{\begingroup \@sanitize@url \@href}%
\providecommand \@href[1]{\@@startlink{#1}\@@href}%
\providecommand \@@href[1]{\endgroup#1\@@endlink}%
\providecommand \@sanitize@url [0]{\catcode `\\12\catcode `\$12\catcode `\&12\catcode `\#12\catcode `\^12\catcode `\_12\catcode `\%12\relax}%
\providecommand \@@startlink[1]{}%
\providecommand \@@endlink[0]{}%
\providecommand \url  [0]{\begingroup\@sanitize@url \@url }%
\providecommand \@url [1]{\endgroup\@href {#1}{\urlprefix }}%
\providecommand \urlprefix  [0]{URL }%
\providecommand \Eprint [0]{\href }%
\providecommand \doibase [0]{http://dx.doi.org/}%
\providecommand \selectlanguage [0]{\@gobble}%
\providecommand \bibinfo  [0]{\@secondoftwo}%
\providecommand \bibfield  [0]{\@secondoftwo}%
\providecommand \translation [1]{[#1]}%
\providecommand \BibitemOpen [0]{}%
\providecommand \bibitemStop [0]{}%
\providecommand \bibitemNoStop [0]{.\EOS\space}%
\providecommand \EOS [0]{\spacefactor3000\relax}%
\providecommand \BibitemShut  [1]{\csname bibitem#1\endcsname}%
\let\auto@bib@innerbib\@empty
%</preamble>
\bibitem [{\citenamefont {Anderson}(1958)}]{Anderson1958}%
  \BibitemOpen
  \bibfield  {author} {\bibinfo {author} {\bibfnamefont {P.~W.}\ \bibnamefont {Anderson}},\ }\bibfield  {title} {\enquote {\bibinfo {title} {Absence of diffusion in certain random lattices},}\ }\href {\doibase 10.1103/PhysRev.109.1492} {\bibfield  {journal} {\bibinfo  {journal} {Phys. Rev.}\ }\textbf {\bibinfo {volume} {109}},\ \bibinfo {pages} {1492--1505} (\bibinfo {year} {1958})}\BibitemShut {NoStop}%
\bibitem [{\citenamefont {Mott}(1987)}]{Mott1987}%
  \BibitemOpen
  \bibfield  {author} {\bibinfo {author} {\bibfnamefont {N.}~\bibnamefont {Mott}},\ }\bibfield  {title} {\enquote {\bibinfo {title} {The mobility edge since 1967},}\ }\href {\doibase 10.1088/0022-3719/20/21/008} {\bibfield  {journal} {\bibinfo  {journal} {J. Phys. C: Solid State Phys.}\ }\textbf {\bibinfo {volume} {20}},\ \bibinfo {pages} {3075} (\bibinfo {year} {1987})}\BibitemShut {NoStop}%
\bibitem [{\citenamefont {Lagendijk}\ \emph {et~al.}(2009)\citenamefont {Lagendijk}, \citenamefont {Tiggelen},\ and\ \citenamefont {Wiersma}}]{Lagendijk2009}%
  \BibitemOpen
  \bibfield  {author} {\bibinfo {author} {\bibfnamefont {A.}~\bibnamefont {Lagendijk}}, \bibinfo {author} {\bibfnamefont {B.~van}\ \bibnamefont {Tiggelen}}, \ and\ \bibinfo {author} {\bibfnamefont {D.~S.}\ \bibnamefont {Wiersma}},\ }\bibfield  {title} {\enquote {\bibinfo {title} {{Fifty years of Anderson localization}},}\ }\href {\doibase 10.1063/1.3206091} {\bibfield  {journal} {\bibinfo  {journal} {Physics Today}\ }\textbf {\bibinfo {volume} {62}},\ \bibinfo {pages} {24--29} (\bibinfo {year} {2009})}\BibitemShut {NoStop}%
\bibitem [{\citenamefont {Abrahams}(2010)}]{AbrahamsBook}%
  \BibitemOpen
  \bibfield  {author} {\bibinfo {author} {\bibfnamefont {E.}~\bibnamefont {Abrahams}},\ }\href@noop {} {\emph {\bibinfo {title} {50 years of Anderson localization}}}\ (\bibinfo  {publisher} {World Scientific},\ \bibinfo {year} {2010})\BibitemShut {NoStop}%
\bibitem [{\citenamefont {Y.~Bliokh}\ and\ \citenamefont {Nori}(2012)}]{Bliokh2012}%
  \BibitemOpen
  \bibfield  {author} {\bibinfo {author} {\bibfnamefont {V.~Freilikher}\ \bibnamefont {Y.~Bliokh}, \bibfnamefont {K.Y.~Bliokh}}\ and\ \bibinfo {author} {\bibfnamefont {F.}~\bibnamefont {Nori}},\ }\enquote {\bibinfo {title} {Anderson localization of light in layered dielectric structures},}\ in\ \href {https://doi.org/10.1201/b12175} {\emph {\bibinfo {booktitle} {Optical Properties of Photonic Structures}}}\ (\bibinfo  {publisher} {CRC Press},\ \bibinfo {year} {2012})\ pp.\ \bibinfo {pages} {55--86}\BibitemShut {NoStop}%
\bibitem [{\citenamefont {I.~M.~Lifshits}(1988)}]{LifshitsBook}%
  \BibitemOpen
  \bibfield  {author} {\bibinfo {author} {\bibfnamefont {L.~A.~Pastur}\ \bibnamefont {I.~M.~Lifshits}, \bibfnamefont {S.~A.~Gredeskul}},\ }\href@noop {} {\emph {\bibinfo {title} {Introduction to the Theory of Disordered Systems}}}\ (\bibinfo  {publisher} {Wiley-VCH},\ \bibinfo {year} {1988})\BibitemShut {NoStop}%
\bibitem [{\citenamefont {Fishman}\ \emph {et~al.}(1982)\citenamefont {Fishman}, \citenamefont {Grempel},\ and\ \citenamefont {Prange}}]{Fishman1982}%
  \BibitemOpen
  \bibfield  {author} {\bibinfo {author} {\bibfnamefont {S.}~\bibnamefont {Fishman}}, \bibinfo {author} {\bibfnamefont {D.~R.}\ \bibnamefont {Grempel}}, \ and\ \bibinfo {author} {\bibfnamefont {R.~E.}\ \bibnamefont {Prange}},\ }\bibfield  {title} {\enquote {\bibinfo {title} {Chaos, quantum recurrences, and {A}nderson localization},}\ }\href {\doibase 10.1103/PhysRevLett.49.509} {\bibfield  {journal} {\bibinfo  {journal} {Phys. Rev. Lett.}\ }\textbf {\bibinfo {volume} {49}},\ \bibinfo {pages} {509--512} (\bibinfo {year} {1982})}\BibitemShut {NoStop}%
\bibitem [{\citenamefont {Aubry}\ and\ \citenamefont {Andr\'e}(1980)}]{Aubry1980}%
  \BibitemOpen
  \bibfield  {author} {\bibinfo {author} {\bibfnamefont {S.}~\bibnamefont {Aubry}}\ and\ \bibinfo {author} {\bibfnamefont {G.}~\bibnamefont {Andr\'e}},\ }\bibfield  {title} {\enquote {\bibinfo {title} {Analyticity breaking and {A}nderson localization in incommensurate lattices},}\ }\href@noop {} {\bibfield  {journal} {\bibinfo  {journal} {Ann. Israel Phys. Soc.}\ }\textbf {\bibinfo {volume} {3}},\ \bibinfo {pages} {18} (\bibinfo {year} {1980})}\BibitemShut {NoStop}%
\bibitem [{\citenamefont {Grempel}\ \emph {et~al.}(1982)\citenamefont {Grempel}, \citenamefont {Fishman},\ and\ \citenamefont {Prange}}]{Grempel1982}%
  \BibitemOpen
  \bibfield  {author} {\bibinfo {author} {\bibfnamefont {D.~R.}\ \bibnamefont {Grempel}}, \bibinfo {author} {\bibfnamefont {Shmuel}\ \bibnamefont {Fishman}}, \ and\ \bibinfo {author} {\bibfnamefont {R.~E.}\ \bibnamefont {Prange}},\ }\bibfield  {title} {\enquote {\bibinfo {title} {Localization in an incommensurate potential: {A}n exactly solvable model},}\ }\href {\doibase 10.1103/PhysRevLett.49.833} {\bibfield  {journal} {\bibinfo  {journal} {Phys. Rev. Lett.}\ }\textbf {\bibinfo {volume} {49}},\ \bibinfo {pages} {833--836} (\bibinfo {year} {1982})}\BibitemShut {NoStop}%
\bibitem [{\citenamefont {Wang}\ \emph {et~al.}(2019)\citenamefont {Wang}, \citenamefont {Zheng}, \citenamefont {Chen} \emph {et~al.}}]{Wang2019_nat}%
  \BibitemOpen
  \bibfield  {author} {\bibinfo {author} {\bibfnamefont {P.}~\bibnamefont {Wang}}, \bibinfo {author} {\bibfnamefont {Y.}~\bibnamefont {Zheng}}, \bibinfo {author} {\bibfnamefont {X.}~\bibnamefont {Chen}},  \emph {et~al.},\ }\bibfield  {title} {\enquote {\bibinfo {title} {Localization and delocalization of light in photonic {M}oir\'e lattices},}\ }\href {\doibase 10.1038/s41586-019-1851-6} {\bibfield  {journal} {\bibinfo  {journal} {Nature}\ }\textbf {\bibinfo {volume} {577}},\ \bibinfo {pages} {42–46} (\bibinfo {year} {2019})}\BibitemShut {NoStop}%
\bibitem [{\citenamefont {Sgrignuoli}\ \emph {et~al.}(2019)\citenamefont {Sgrignuoli}, \citenamefont {Wang}, \citenamefont {Pinheiro},\ and\ \citenamefont {Dal~Negro}}]{Sprignuoli2019}%
  \BibitemOpen
  \bibfield  {author} {\bibinfo {author} {\bibfnamefont {F.}~\bibnamefont {Sgrignuoli}}, \bibinfo {author} {\bibfnamefont {R.}~\bibnamefont {Wang}}, \bibinfo {author} {\bibfnamefont {F.~A.}\ \bibnamefont {Pinheiro}}, \ and\ \bibinfo {author} {\bibfnamefont {L.}~\bibnamefont {Dal~Negro}},\ }\bibfield  {title} {\enquote {\bibinfo {title} {Localization of scattering resonances in aperiodic {V}ogel spirals},}\ }\href {\doibase 10.1103/PhysRevB.99.104202} {\bibfield  {journal} {\bibinfo  {journal} {Phys. Rev. B}\ }\textbf {\bibinfo {volume} {99}},\ \bibinfo {pages} {104202} (\bibinfo {year} {2019})}\BibitemShut {NoStop}%
\bibitem [{\citenamefont {Zhu}\ \emph {et~al.}(2024)\citenamefont {Zhu}, \citenamefont {Yu}, \citenamefont {Johnstone},\ and\ \citenamefont {Sanchez-Palencia}}]{Zhu2024}%
  \BibitemOpen
  \bibfield  {author} {\bibinfo {author} {\bibfnamefont {Z.}~\bibnamefont {Zhu}}, \bibinfo {author} {\bibfnamefont {S.}~\bibnamefont {Yu}}, \bibinfo {author} {\bibfnamefont {D.}~\bibnamefont {Johnstone}}, \ and\ \bibinfo {author} {\bibfnamefont {L.}~\bibnamefont {Sanchez-Palencia}},\ }\bibfield  {title} {\enquote {\bibinfo {title} {Localization and spectral structure in two-dimensional quasicrystal potentials},}\ }\href {\doibase 10.1103/PhysRevA.109.013314} {\bibfield  {journal} {\bibinfo  {journal} {Phys. Rev. A}\ }\textbf {\bibinfo {volume} {109}},\ \bibinfo {pages} {013314} (\bibinfo {year} {2024})}\BibitemShut {NoStop}%
\bibitem [{\citenamefont {Freedman}\ \emph {et~al.}(2006)\citenamefont {Freedman}, \citenamefont {Bartal}, \citenamefont {Segev} \emph {et~al.}}]{Freedman2006}%
  \BibitemOpen
  \bibfield  {author} {\bibinfo {author} {\bibfnamefont {B.}~\bibnamefont {Freedman}}, \bibinfo {author} {\bibfnamefont {G.}~\bibnamefont {Bartal}}, \bibinfo {author} {\bibfnamefont {M.}~\bibnamefont {Segev}},  \emph {et~al.},\ }\bibfield  {title} {\enquote {\bibinfo {title} {Wave and defect dynamics in nonlinear photonic quasicrystals},}\ }\href {\doibase 10.1038/nature04722} {\bibfield  {journal} {\bibinfo  {journal} {Nature}\ }\textbf {\bibinfo {volume} {440}},\ \bibinfo {pages} {1166–1169} (\bibinfo {year} {2006})}\BibitemShut {NoStop}%
\bibitem [{\citenamefont {Greiner}\ \emph {et~al.}(2002)\citenamefont {Greiner}, \citenamefont {Mandel}, \citenamefont {Esslinger}, \citenamefont {H\"{a}nsch},\ and\ \citenamefont {Bloch}}]{Greiner2002}%
  \BibitemOpen
  \bibfield  {author} {\bibinfo {author} {\bibfnamefont {Markus}\ \bibnamefont {Greiner}}, \bibinfo {author} {\bibfnamefont {Olaf}\ \bibnamefont {Mandel}}, \bibinfo {author} {\bibfnamefont {Tilman}\ \bibnamefont {Esslinger}}, \bibinfo {author} {\bibfnamefont {Theodor~W.}\ \bibnamefont {H\"{a}nsch}}, \ and\ \bibinfo {author} {\bibfnamefont {Immanuel}\ \bibnamefont {Bloch}},\ }\bibfield  {title} {\enquote {\bibinfo {title} {Quantum phase transition from a superfluid to a {M}ott insulator in a gas of ultracold atoms},}\ }\href {\doibase 10.1038/415039a} {\bibfield  {journal} {\bibinfo  {journal} {Nature}\ }\textbf {\bibinfo {volume} {415}},\ \bibinfo {pages} {39–44} (\bibinfo {year} {2002})}\BibitemShut {NoStop}%
\bibitem [{\citenamefont {J\"{o}rdens}\ \emph {et~al.}(2008)\citenamefont {J\"{o}rdens}, \citenamefont {Strohmaier}, \citenamefont {G\"{u}nter}, \citenamefont {Moritz},\ and\ \citenamefont {Esslinger}}]{Jordens2008}%
  \BibitemOpen
  \bibfield  {author} {\bibinfo {author} {\bibfnamefont {Robert}\ \bibnamefont {J\"{o}rdens}}, \bibinfo {author} {\bibfnamefont {Niels}\ \bibnamefont {Strohmaier}}, \bibinfo {author} {\bibfnamefont {Kenneth}\ \bibnamefont {G\"{u}nter}}, \bibinfo {author} {\bibfnamefont {Henning}\ \bibnamefont {Moritz}}, \ and\ \bibinfo {author} {\bibfnamefont {Tilman}\ \bibnamefont {Esslinger}},\ }\bibfield  {title} {\enquote {\bibinfo {title} {A {M}ott insulator of fermionic atoms in an optical lattice},}\ }\href {\doibase 10.1038/nature07244} {\bibfield  {journal} {\bibinfo  {journal} {Nature}\ }\textbf {\bibinfo {volume} {455}},\ \bibinfo {pages} {204–207} (\bibinfo {year} {2008})}\BibitemShut {NoStop}%
\bibitem [{\citenamefont {Smith}\ \emph {et~al.}(2017)\citenamefont {Smith}, \citenamefont {Knolle}, \citenamefont {Kovrizhin},\ and\ \citenamefont {Moessner}}]{Smith2017}%
  \BibitemOpen
  \bibfield  {author} {\bibinfo {author} {\bibfnamefont {A.}~\bibnamefont {Smith}}, \bibinfo {author} {\bibfnamefont {J.}~\bibnamefont {Knolle}}, \bibinfo {author} {\bibfnamefont {D.~L.}\ \bibnamefont {Kovrizhin}}, \ and\ \bibinfo {author} {\bibfnamefont {R.}~\bibnamefont {Moessner}},\ }\bibfield  {title} {\enquote {\bibinfo {title} {Disorder-free localization},}\ }\href {\doibase 10.1103/PhysRevLett.118.266601} {\bibfield  {journal} {\bibinfo  {journal} {Phys. Rev. Lett.}\ }\textbf {\bibinfo {volume} {118}},\ \bibinfo {pages} {266601} (\bibinfo {year} {2017})}\BibitemShut {NoStop}%
\bibitem [{\citenamefont {Brenes}\ \emph {et~al.}(2018)\citenamefont {Brenes}, \citenamefont {Dalmonte}, \citenamefont {Heyl},\ and\ \citenamefont {Scardicchio}}]{Brenes2018}%
  \BibitemOpen
  \bibfield  {author} {\bibinfo {author} {\bibfnamefont {M.}~\bibnamefont {Brenes}}, \bibinfo {author} {\bibfnamefont {M.}~\bibnamefont {Dalmonte}}, \bibinfo {author} {\bibfnamefont {M.}~\bibnamefont {Heyl}}, \ and\ \bibinfo {author} {\bibfnamefont {A.}~\bibnamefont {Scardicchio}},\ }\bibfield  {title} {\enquote {\bibinfo {title} {Many-body localization dynamics from gauge invariance},}\ }\href {\doibase 10.1103/PhysRevLett.120.030601} {\bibfield  {journal} {\bibinfo  {journal} {Phys. Rev. Lett.}\ }\textbf {\bibinfo {volume} {120}},\ \bibinfo {pages} {030601} (\bibinfo {year} {2018})}\BibitemShut {NoStop}%
\bibitem [{\citenamefont {Chakraborty}\ \emph {et~al.}(2022)\citenamefont {Chakraborty}, \citenamefont {Heyl}, \citenamefont {Karpov},\ and\ \citenamefont {Moessner}}]{Chakraborty2022}%
  \BibitemOpen
  \bibfield  {author} {\bibinfo {author} {\bibfnamefont {N.}~\bibnamefont {Chakraborty}}, \bibinfo {author} {\bibfnamefont {M.}~\bibnamefont {Heyl}}, \bibinfo {author} {\bibfnamefont {P.}~\bibnamefont {Karpov}}, \ and\ \bibinfo {author} {\bibfnamefont {R.}~\bibnamefont {Moessner}},\ }\bibfield  {title} {\enquote {\bibinfo {title} {Disorder-free localization transition in a two-dimensional lattice gauge theory},}\ }\href {\doibase 10.1103/PhysRevB.106.L060308} {\bibfield  {journal} {\bibinfo  {journal} {Phys. Rev. B}\ }\textbf {\bibinfo {volume} {106}},\ \bibinfo {pages} {L060308} (\bibinfo {year} {2022})}\BibitemShut {NoStop}%
\bibitem [{\citenamefont {Vicencio}(2021)}]{Vicencio2021}%
  \BibitemOpen
  \bibfield  {author} {\bibinfo {author} {\bibfnamefont {R.~A.}\ \bibnamefont {Vicencio}},\ }\bibfield  {title} {\enquote {\bibinfo {title} {Photonic flat band dynamics},}\ }\href {\doibase 10.1080/23746149.2021.1878057} {\bibfield  {journal} {\bibinfo  {journal} {Adv. Phys. X}\ }\textbf {\bibinfo {volume} {6}},\ \bibinfo {pages} {1878057} (\bibinfo {year} {2021})}\BibitemShut {NoStop}%
\bibitem [{\citenamefont {Danieli}\ \emph {et~al.}(2024)\citenamefont {Danieli}, \citenamefont {Andreanov}, \citenamefont {Leykam},\ and\ \citenamefont {Flach}}]{Danieli2024}%
  \BibitemOpen
  \bibfield  {author} {\bibinfo {author} {\bibfnamefont {C.}~\bibnamefont {Danieli}}, \bibinfo {author} {\bibfnamefont {A.}~\bibnamefont {Andreanov}}, \bibinfo {author} {\bibfnamefont {D.}~\bibnamefont {Leykam}}, \ and\ \bibinfo {author} {\bibfnamefont {S.}~\bibnamefont {Flach}},\ }\bibfield  {title} {\enquote {\bibinfo {title} {Flat band fine-tuning and its photonic applications},}\ }\href {\doibase doi:10.1515/nanoph-2024-0135} {\bibfield  {journal} {\bibinfo  {journal} {Nanophotonics}\ }\textbf {\bibinfo {volume} {13}},\ \bibinfo {pages} {3925--3944} (\bibinfo {year} {2024})}\BibitemShut {NoStop}%
\bibitem [{\citenamefont {Nori}\ and\ \citenamefont {Niu}(1990)}]{Nori1990}%
  \BibitemOpen
  \bibfield  {author} {\bibinfo {author} {\bibfnamefont {F.}~\bibnamefont {Nori}}\ and\ \bibinfo {author} {\bibfnamefont {Q.}~\bibnamefont {Niu}},\ }\bibfield  {title} {\enquote {\bibinfo {title} {Angular momentum irreducible representation and destructive quantum interference for {P}enrose lattice {H}amiltonians},}\ }\href@noop {} {\bibfield  {journal} {\bibinfo  {journal} {Quasi Crystals and Incommensurate Structures in Condensed Matter (World Scientific, Singapore, 1990)}\ ,\ \bibinfo {pages} {434}} (\bibinfo {year} {1990})}\BibitemShut {NoStop}%
\bibitem [{\citenamefont {Flach}\ \emph {et~al.}(2014)\citenamefont {Flach}, \citenamefont {Leykam}, \citenamefont {Bodyfelt}, \citenamefont {Matthies},\ and\ \citenamefont {Desyatnikov}}]{Flach2014}%
  \BibitemOpen
  \bibfield  {author} {\bibinfo {author} {\bibfnamefont {S.}~\bibnamefont {Flach}}, \bibinfo {author} {\bibfnamefont {D.}~\bibnamefont {Leykam}}, \bibinfo {author} {\bibfnamefont {J.~D.}\ \bibnamefont {Bodyfelt}}, \bibinfo {author} {\bibfnamefont {P.}~\bibnamefont {Matthies}}, \ and\ \bibinfo {author} {\bibfnamefont {A.~S.}\ \bibnamefont {Desyatnikov}},\ }\bibfield  {title} {\enquote {\bibinfo {title} {Detangling flat bands into {F}ano lattices},}\ }\href {\doibase 10.1209/0295-5075/105/30001} {\bibfield  {journal} {\bibinfo  {journal} {Europhys. Lett.}\ }\textbf {\bibinfo {volume} {105}},\ \bibinfo {pages} {30001} (\bibinfo {year} {2014})}\BibitemShut {NoStop}%
\bibitem [{\citenamefont {Maimaiti}\ \emph {et~al.}(2017)\citenamefont {Maimaiti}, \citenamefont {Andreanov}, \citenamefont {Park}, \citenamefont {Gendelman},\ and\ \citenamefont {Flach}}]{Maimaiti2017}%
  \BibitemOpen
  \bibfield  {author} {\bibinfo {author} {\bibfnamefont {W.}~\bibnamefont {Maimaiti}}, \bibinfo {author} {\bibfnamefont {A.}~\bibnamefont {Andreanov}}, \bibinfo {author} {\bibfnamefont {H.~C.}\ \bibnamefont {Park}}, \bibinfo {author} {\bibfnamefont {O.}~\bibnamefont {Gendelman}}, \ and\ \bibinfo {author} {\bibfnamefont {S.}~\bibnamefont {Flach}},\ }\bibfield  {title} {\enquote {\bibinfo {title} {Compact localized states and flat-band generators in one dimension},}\ }\href {\doibase 10.1103/PhysRevB.95.115135} {\bibfield  {journal} {\bibinfo  {journal} {Phys. Rev. B}\ }\textbf {\bibinfo {volume} {95}},\ \bibinfo {pages} {115135} (\bibinfo {year} {2017})}\BibitemShut {NoStop}%
\bibitem [{\citenamefont {R\"ontgen}\ \emph {et~al.}(2019)\citenamefont {R\"ontgen}, \citenamefont {Morfonios}, \citenamefont {Brouzos}, \citenamefont {Diakonos},\ and\ \citenamefont {Schmelcher}}]{Rontgen2018let}%
  \BibitemOpen
  \bibfield  {author} {\bibinfo {author} {\bibfnamefont {M.}~\bibnamefont {R\"ontgen}}, \bibinfo {author} {\bibfnamefont {C.~V.}\ \bibnamefont {Morfonios}}, \bibinfo {author} {\bibfnamefont {I.}~\bibnamefont {Brouzos}}, \bibinfo {author} {\bibfnamefont {F.~K.}\ \bibnamefont {Diakonos}}, \ and\ \bibinfo {author} {\bibfnamefont {P.}~\bibnamefont {Schmelcher}},\ }\bibfield  {title} {\enquote {\bibinfo {title} {Quantum network transfer and storage with compact localized states induced by local symmetries},}\ }\href {\doibase 10.1103/PhysRevLett.123.080504} {\bibfield  {journal} {\bibinfo  {journal} {Phys. Rev. Lett.}\ }\textbf {\bibinfo {volume} {123}},\ \bibinfo {pages} {080504} (\bibinfo {year} {2019})}\BibitemShut {NoStop}%
\bibitem [{\citenamefont {R\"ontgen}\ \emph {et~al.}(2018)\citenamefont {R\"ontgen}, \citenamefont {Morfonios},\ and\ \citenamefont {Schmelcher}}]{Rontgen2018}%
  \BibitemOpen
  \bibfield  {author} {\bibinfo {author} {\bibfnamefont {M.}~\bibnamefont {R\"ontgen}}, \bibinfo {author} {\bibfnamefont {C.~V.}\ \bibnamefont {Morfonios}}, \ and\ \bibinfo {author} {\bibfnamefont {P.}~\bibnamefont {Schmelcher}},\ }\bibfield  {title} {\enquote {\bibinfo {title} {Compact localized states and flat bands from local symmetry partitioning},}\ }\href {\doibase 10.1103/PhysRevB.97.035161} {\bibfield  {journal} {\bibinfo  {journal} {Phys. Rev. B}\ }\textbf {\bibinfo {volume} {97}},\ \bibinfo {pages} {035161} (\bibinfo {year} {2018})}\BibitemShut {NoStop}%
\bibitem [{\citenamefont {D.~Leykam}\ and\ \citenamefont {Flach}(2018)}]{Leykam2018}%
  \BibitemOpen
  \bibfield  {author} {\bibinfo {author} {\bibfnamefont {A.~Andreanov}\ \bibnamefont {D.~Leykam}}\ and\ \bibinfo {author} {\bibfnamefont {S.}~\bibnamefont {Flach}},\ }\bibfield  {title} {\enquote {\bibinfo {title} {Artificial flat band systems: from lattice models to experiments},}\ }\href {\doibase 10.1080/23746149.2018.1473052} {\bibfield  {journal} {\bibinfo  {journal} {Adv. Phys.: X}\ }\textbf {\bibinfo {volume} {3}},\ \bibinfo {pages} {1473052} (\bibinfo {year} {2018})}\BibitemShut {NoStop}%
\bibitem [{\citenamefont {Leykam}\ and\ \citenamefont {Flach}(2018)}]{Leykam2018b}%
  \BibitemOpen
  \bibfield  {author} {\bibinfo {author} {\bibfnamefont {D.}~\bibnamefont {Leykam}}\ and\ \bibinfo {author} {\bibfnamefont {S.}~\bibnamefont {Flach}},\ }\bibfield  {title} {\enquote {\bibinfo {title} {Perspective: {P}hotonic flatbands},}\ }\href {\doibase 10.1063/1.5034365} {\bibfield  {journal} {\bibinfo  {journal} {APL Photon.}\ }\textbf {\bibinfo {volume} {3}},\ \bibinfo {pages} {070901} (\bibinfo {year} {2018})}\BibitemShut {NoStop}%
\bibitem [{\citenamefont {Danieli}\ \emph {et~al.}(2020)\citenamefont {Danieli}, \citenamefont {Andreanov},\ and\ \citenamefont {Flach}}]{Danieli2020}%
  \BibitemOpen
  \bibfield  {author} {\bibinfo {author} {\bibfnamefont {C.}~\bibnamefont {Danieli}}, \bibinfo {author} {\bibfnamefont {A.}~\bibnamefont {Andreanov}}, \ and\ \bibinfo {author} {\bibfnamefont {S.}~\bibnamefont {Flach}},\ }\bibfield  {title} {\enquote {\bibinfo {title} {Many-body flatband localization},}\ }\href {\doibase 10.1103/PhysRevB.102.041116} {\bibfield  {journal} {\bibinfo  {journal} {Phys. Rev. B}\ }\textbf {\bibinfo {volume} {102}},\ \bibinfo {pages} {041116} (\bibinfo {year} {2020})}\BibitemShut {NoStop}%
\bibitem [{\citenamefont {Hatano}\ and\ \citenamefont {Nelson}(1996)}]{Hatano1996}%
  \BibitemOpen
  \bibfield  {author} {\bibinfo {author} {\bibfnamefont {N.}~\bibnamefont {Hatano}}\ and\ \bibinfo {author} {\bibfnamefont {D.~R.}\ \bibnamefont {Nelson}},\ }\bibfield  {title} {\enquote {\bibinfo {title} {Localization transitions in non-{H}ermitian quantum mechanics},}\ }\href {\doibase 10.1103/PhysRevLett.77.570} {\bibfield  {journal} {\bibinfo  {journal} {Phys. Rev. Lett.}\ }\textbf {\bibinfo {volume} {77}},\ \bibinfo {pages} {570--573} (\bibinfo {year} {1996})}\BibitemShut {NoStop}%
\bibitem [{\citenamefont {Hatano}\ and\ \citenamefont {Nelson}(1997)}]{Hatano1997}%
  \BibitemOpen
  \bibfield  {author} {\bibinfo {author} {\bibfnamefont {N.}~\bibnamefont {Hatano}}\ and\ \bibinfo {author} {\bibfnamefont {D.~R.}\ \bibnamefont {Nelson}},\ }\bibfield  {title} {\enquote {\bibinfo {title} {Vortex pinning and non-{H}ermitian quantum mechanics},}\ }\href {\doibase 10.1103/PhysRevB.56.8651} {\bibfield  {journal} {\bibinfo  {journal} {Phys. Rev. B}\ }\textbf {\bibinfo {volume} {56}},\ \bibinfo {pages} {8651--8673} (\bibinfo {year} {1997})}\BibitemShut {NoStop}%
\bibitem [{\citenamefont {Yao}\ and\ \citenamefont {Wang}(2018)}]{Yao2018}%
  \BibitemOpen
  \bibfield  {author} {\bibinfo {author} {\bibfnamefont {S.}~\bibnamefont {Yao}}\ and\ \bibinfo {author} {\bibfnamefont {Z.}~\bibnamefont {Wang}},\ }\bibfield  {title} {\enquote {\bibinfo {title} {Edge states and topological invariants of non-{H}ermitian systems},}\ }\href {\doibase 10.1103/PhysRevLett.121.086803} {\bibfield  {journal} {\bibinfo  {journal} {Phys. Rev. Lett.}\ }\textbf {\bibinfo {volume} {121}},\ \bibinfo {pages} {086803} (\bibinfo {year} {2018})}\BibitemShut {NoStop}%
\bibitem [{\citenamefont {Martinez~Alvarez}\ \emph {et~al.}(2018)\citenamefont {Martinez~Alvarez}, \citenamefont {Barrios~Vargas},\ and\ \citenamefont {Foa~Torres}}]{Alvarez2018}%
  \BibitemOpen
  \bibfield  {author} {\bibinfo {author} {\bibfnamefont {V.~M.}\ \bibnamefont {Martinez~Alvarez}}, \bibinfo {author} {\bibfnamefont {J.~E.}\ \bibnamefont {Barrios~Vargas}}, \ and\ \bibinfo {author} {\bibfnamefont {L.~E.~F.}\ \bibnamefont {Foa~Torres}},\ }\bibfield  {title} {\enquote {\bibinfo {title} {Non-{H}ermitian robust edge states in one dimension: {A}nomalous localization and eigenspace condensation at exceptional points},}\ }\href {\doibase 10.1103/PhysRevB.97.121401} {\bibfield  {journal} {\bibinfo  {journal} {Phys. Rev. B}\ }\textbf {\bibinfo {volume} {97}},\ \bibinfo {pages} {121401} (\bibinfo {year} {2018})}\BibitemShut {NoStop}%
\bibitem [{\citenamefont {Arkhipov}\ and\ \citenamefont {Minganti}(2023)}]{arkhipov2023a}%
  \BibitemOpen
  \bibfield  {author} {\bibinfo {author} {\bibfnamefont {I.~I.}\ \bibnamefont {Arkhipov}}\ and\ \bibinfo {author} {\bibfnamefont {F.}~\bibnamefont {Minganti}},\ }\bibfield  {title} {\enquote {\bibinfo {title} {Emergent non-{H}ermitian localization phenomena in the synthetic space of zero-dimensional bosonic systems},}\ }\href {\doibase 10.1103/PhysRevA.107.012202} {\bibfield  {journal} {\bibinfo  {journal} {Phys. Rev. A}\ }\textbf {\bibinfo {volume} {107}},\ \bibinfo {pages} {012202} (\bibinfo {year} {2023})}\BibitemShut {NoStop}%
\bibitem [{\citenamefont {Ozawa}\ \emph {et~al.}(2019)\citenamefont {Ozawa}, \citenamefont {Price}, \citenamefont {Amo}, \citenamefont {Goldman}, \citenamefont {Hafezi}, \citenamefont {Lu}, \citenamefont {Rechtsman}, \citenamefont {Schuster}, \citenamefont {Simon}, \citenamefont {Zilberberg},\ and\ \citenamefont {Carusotto}}]{Ozawa2019}%
  \BibitemOpen
  \bibfield  {author} {\bibinfo {author} {\bibfnamefont {T.}~\bibnamefont {Ozawa}}, \bibinfo {author} {\bibfnamefont {H.~M.}\ \bibnamefont {Price}}, \bibinfo {author} {\bibfnamefont {A.}~\bibnamefont {Amo}}, \bibinfo {author} {\bibfnamefont {N.}~\bibnamefont {Goldman}}, \bibinfo {author} {\bibfnamefont {M.}~\bibnamefont {Hafezi}}, \bibinfo {author} {\bibfnamefont {L.}~\bibnamefont {Lu}}, \bibinfo {author} {\bibfnamefont {M.~C.}\ \bibnamefont {Rechtsman}}, \bibinfo {author} {\bibfnamefont {D.}~\bibnamefont {Schuster}}, \bibinfo {author} {\bibfnamefont {J.}~\bibnamefont {Simon}}, \bibinfo {author} {\bibfnamefont {O.}~\bibnamefont {Zilberberg}}, \ and\ \bibinfo {author} {\bibfnamefont {I.}~\bibnamefont {Carusotto}},\ }\bibfield  {title} {\enquote {\bibinfo {title} {Topological photonics},}\ }\href {\doibase 10.1103/RevModPhys.91.015006} {\bibfield  {journal} {\bibinfo  {journal} {Rev. Mod. Phys.}\ }\textbf {\bibinfo {volume} {91}},\ \bibinfo {pages} {015006} (\bibinfo {year} {2019})}\BibitemShut {NoStop}%
\bibitem [{\citenamefont {Bergholtz}\ \emph {et~al.}(2021)\citenamefont {Bergholtz}, \citenamefont {Budich},\ and\ \citenamefont {Kunst}}]{Bergholtz2021}%
  \BibitemOpen
  \bibfield  {author} {\bibinfo {author} {\bibfnamefont {E.~J.}\ \bibnamefont {Bergholtz}}, \bibinfo {author} {\bibfnamefont {J.~C.}\ \bibnamefont {Budich}}, \ and\ \bibinfo {author} {\bibfnamefont {F.~K.}\ \bibnamefont {Kunst}},\ }\bibfield  {title} {\enquote {\bibinfo {title} {Exceptional topology of non-{H}ermitian systems},}\ }\href {\doibase 10.1103/RevModPhys.93.015005} {\bibfield  {journal} {\bibinfo  {journal} {Rev. Mod. Phys.}\ }\textbf {\bibinfo {volume} {93}},\ \bibinfo {pages} {015005} (\bibinfo {year} {2021})}\BibitemShut {NoStop}%
\bibitem [{\citenamefont {Weimann}\ \emph {et~al.}(2017)\citenamefont {Weimann}, \citenamefont {Kremer}, \citenamefont {Plotnik}, \citenamefont {Lumer}, \citenamefont {Nolte}, \citenamefont {Makris}, \citenamefont {Segev}, \citenamefont {Rechtsman},\ and\ \citenamefont {Szameit}}]{Weimann2017}%
  \BibitemOpen
  \bibfield  {author} {\bibinfo {author} {\bibfnamefont {S.}~\bibnamefont {Weimann}}, \bibinfo {author} {\bibfnamefont {M.}~\bibnamefont {Kremer}}, \bibinfo {author} {\bibfnamefont {Y.}~\bibnamefont {Plotnik}}, \bibinfo {author} {\bibfnamefont {Y.}~\bibnamefont {Lumer}}, \bibinfo {author} {\bibfnamefont {S.}~\bibnamefont {Nolte}}, \bibinfo {author} {\bibfnamefont {{K. G.}}\ \bibnamefont {Makris}}, \bibinfo {author} {\bibfnamefont {M.}~\bibnamefont {Segev}}, \bibinfo {author} {\bibfnamefont {{M. C.}}\ \bibnamefont {Rechtsman}}, \ and\ \bibinfo {author} {\bibfnamefont {A.}~\bibnamefont {Szameit}},\ }\bibfield  {title} {\enquote {\bibinfo {title} {Topologically protected bound states in photonic parity–time-symmetric crystals},}\ }\href {\doibase 10.1038/NMAT4811} {\bibfield  {journal} {\bibinfo  {journal} {Nat. Mater.}\ }\textbf {\bibinfo {volume} {16}},\ \bibinfo {pages} {433--438} (\bibinfo {year} {2017})}\BibitemShut {NoStop}%
\bibitem [{\citenamefont {Roy}\ \emph {et~al.}(2022)\citenamefont {Roy}, \citenamefont {Parto}, \citenamefont {Nehra}, \citenamefont {Leefmans},\ and\ \citenamefont {Marandi}}]{Roy2021b}%
  \BibitemOpen
  \bibfield  {author} {\bibinfo {author} {\bibfnamefont {A.}~\bibnamefont {Roy}}, \bibinfo {author} {\bibfnamefont {M.}~\bibnamefont {Parto}}, \bibinfo {author} {\bibfnamefont {R.}~\bibnamefont {Nehra}}, \bibinfo {author} {\bibfnamefont {C.}~\bibnamefont {Leefmans}}, \ and\ \bibinfo {author} {\bibfnamefont {A.}~\bibnamefont {Marandi}},\ }\bibfield  {title} {\enquote {\bibinfo {title} {Topological optical parametric oscillation},}\ }\href {\doibase doi:10.1515/nanoph-2021-0765} {\bibfield  {journal} {\bibinfo  {journal} {Nanophotonics}\ }\textbf {\bibinfo {volume} {11}},\ \bibinfo {pages} {1611--1618} (\bibinfo {year} {2022})}\BibitemShut {NoStop}%
\bibitem [{\citenamefont {Weidemann}\ \emph {et~al.}(2020)\citenamefont {Weidemann}, \citenamefont {Kremer}, \citenamefont {Helbig}, \citenamefont {Hofmann}, \citenamefont {Stegmaier}, \citenamefont {Greiter}, \citenamefont {Thomale},\ and\ \citenamefont {Szameit}}]{Weidemann2020}%
  \BibitemOpen
  \bibfield  {author} {\bibinfo {author} {\bibfnamefont {S.}~\bibnamefont {Weidemann}}, \bibinfo {author} {\bibfnamefont {M.}~\bibnamefont {Kremer}}, \bibinfo {author} {\bibfnamefont {T.}~\bibnamefont {Helbig}}, \bibinfo {author} {\bibfnamefont {T.}~\bibnamefont {Hofmann}}, \bibinfo {author} {\bibfnamefont {A.}~\bibnamefont {Stegmaier}}, \bibinfo {author} {\bibfnamefont {M.}~\bibnamefont {Greiter}}, \bibinfo {author} {\bibfnamefont {R.}~\bibnamefont {Thomale}}, \ and\ \bibinfo {author} {\bibfnamefont {A.}~\bibnamefont {Szameit}},\ }\bibfield  {title} {\enquote {\bibinfo {title} {Topological funneling of light},}\ }\href {\doibase 10.1126/science.aaz8727} {\bibfield  {journal} {\bibinfo  {journal} {Science}\ }\textbf {\bibinfo {volume} {368}},\ \bibinfo {pages} {311--314} (\bibinfo {year} {2020})}\BibitemShut {NoStop}%
\bibitem [{\citenamefont {K\"{o}nig}(2016)}]{Konig206}%
  \BibitemOpen
  \bibfield  {author} {\bibinfo {author} {\bibfnamefont {W.}~\bibnamefont {K\"{o}nig}},\ }\href@noop {} {\emph {\bibinfo {title} {The Parabolic {A}nderson model, Random Walk in Random Potential}}}\ (\bibinfo  {publisher} {Birkh\"auser},\ \bibinfo {address} {Basel},\ \bibinfo {year} {2016})\BibitemShut {NoStop}%
\bibitem [{\citenamefont {Semenov}\ \emph {et~al.}(2024)\citenamefont {Semenov}, \citenamefont {Murphy}, \citenamefont {Patscheider}, \citenamefont {Bernardi},\ and\ \citenamefont {Blokhina}}]{Semenov2024}%
  \BibitemOpen
  \bibfield  {author} {\bibinfo {author} {\bibfnamefont {A.}~\bibnamefont {Semenov}}, \bibinfo {author} {\bibfnamefont {N.}~\bibnamefont {Murphy}}, \bibinfo {author} {\bibfnamefont {S.}~\bibnamefont {Patscheider}}, \bibinfo {author} {\bibfnamefont {A.}~\bibnamefont {Bernardi}}, \ and\ \bibinfo {author} {\bibfnamefont {E.}~\bibnamefont {Blokhina}},\ }\href@noop {} {\enquote {\bibinfo {title} {A quantum implementation of high-order power method for estimating geometric entanglement of pure states},}\ } (\bibinfo {year} {2024}),\ \Eprint {http://arxiv.org/abs/2405.19134} {arXiv:2405.19134} \BibitemShut {NoStop}%
\bibitem [{\citenamefont {Howard}\ \emph {et~al.}(2019)\citenamefont {Howard}, \citenamefont {Weinhold}, \citenamefont {Shahandeh}, \citenamefont {Combes}, \citenamefont {Vanner}, \citenamefont {White},\ and\ \citenamefont {Ringbauer}}]{Howard2019}%
  \BibitemOpen
  \bibfield  {author} {\bibinfo {author} {\bibfnamefont {L.~A.}\ \bibnamefont {Howard}}, \bibinfo {author} {\bibfnamefont {T.~J.}\ \bibnamefont {Weinhold}}, \bibinfo {author} {\bibfnamefont {F.}~\bibnamefont {Shahandeh}}, \bibinfo {author} {\bibfnamefont {J.}~\bibnamefont {Combes}}, \bibinfo {author} {\bibfnamefont {M.~R.}\ \bibnamefont {Vanner}}, \bibinfo {author} {\bibfnamefont {A.~G.}\ \bibnamefont {White}}, \ and\ \bibinfo {author} {\bibfnamefont {M.}~\bibnamefont {Ringbauer}},\ }\bibfield  {title} {\enquote {\bibinfo {title} {Quantum hypercube states},}\ }\href {\doibase 10.1103/PhysRevLett.123.020402} {\bibfield  {journal} {\bibinfo  {journal} {Phys. Rev. Lett.}\ }\textbf {\bibinfo {volume} {123}},\ \bibinfo {pages} {020402} (\bibinfo {year} {2019})}\BibitemShut {NoStop}%
\bibitem [{\citenamefont {Goto}(2024)}]{Goto2024}%
  \BibitemOpen
  \bibfield  {author} {\bibinfo {author} {\bibfnamefont {H.}~\bibnamefont {Goto}},\ }\bibfield  {title} {\enquote {\bibinfo {title} {High-performance fault-tolerant quantum computing with many-hypercube codes},}\ }\href {\doibase 10.1126/sciadv.adp6388} {\bibfield  {journal} {\bibinfo  {journal} {Sci. Adv.}\ }\textbf {\bibinfo {volume} {10}},\ \bibinfo {pages} {eadp6388} (\bibinfo {year} {2024})}\BibitemShut {NoStop}%
\bibitem [{\citenamefont {Parisi}(1994)}]{Parisi1994}%
  \BibitemOpen
  \bibfield  {author} {\bibinfo {author} {\bibfnamefont {G}~\bibnamefont {Parisi}},\ }\bibfield  {title} {\enquote {\bibinfo {title} {D-dimensional arrays of {J}osephson junctions, spin glasses and q-deformed harmonic oscillators},}\ }\href {\doibase 10.1088/0305-4470/27/23/007} {\bibfield  {journal} {\bibinfo  {journal} {J. Phys. A: Math. Gen.}\ }\textbf {\bibinfo {volume} {27}},\ \bibinfo {pages} {7555} (\bibinfo {year} {1994})}\BibitemShut {NoStop}%
\bibitem [{\citenamefont {Marinari}\ \emph {et~al.}(1995)\citenamefont {Marinari}, \citenamefont {Parisi},\ and\ \citenamefont {Ritort}}]{Marinari1995}%
  \BibitemOpen
  \bibfield  {author} {\bibinfo {author} {\bibfnamefont {E}~\bibnamefont {Marinari}}, \bibinfo {author} {\bibfnamefont {G}~\bibnamefont {Parisi}}, \ and\ \bibinfo {author} {\bibfnamefont {F}~\bibnamefont {Ritort}},\ }\bibfield  {title} {\enquote {\bibinfo {title} {Replica theory and large-d {J}osephson junction hypercubic models},}\ }\href {\doibase 10.1088/0305-4470/28/16/008} {\bibfield  {journal} {\bibinfo  {journal} {J. Phys. A: Math. Gen.}\ }\textbf {\bibinfo {volume} {28}},\ \bibinfo {pages} {4481} (\bibinfo {year} {1995})}\BibitemShut {NoStop}%
\bibitem [{\citenamefont {Jia}\ and\ \citenamefont {Verbaarschot}(2020)}]{Jia2020}%
  \BibitemOpen
  \bibfield  {author} {\bibinfo {author} {\bibfnamefont {Y.}~\bibnamefont {Jia}}\ and\ \bibinfo {author} {\bibfnamefont {J.J.M.}\ \bibnamefont {Verbaarschot}},\ }\bibfield  {title} {\enquote {\bibinfo {title} {Chaos on the hypercube},}\ }\href {\doibase 10.1007/JHEP11(2020)154} {\bibfield  {journal} {\bibinfo  {journal} {J. High Energ. Phys.}\ }\textbf {\bibinfo {volume} {2020}},\ \bibinfo {pages} {154} (\bibinfo {year} {2020})}\BibitemShut {NoStop}%
\bibitem [{\citenamefont {Avena}\ \emph {et~al.}(2020)\citenamefont {Avena}, \citenamefont {Gün},\ and\ \citenamefont {Hesse}}]{Avena2020}%
  \BibitemOpen
  \bibfield  {author} {\bibinfo {author} {\bibfnamefont {L.}~\bibnamefont {Avena}}, \bibinfo {author} {\bibfnamefont {O.}~\bibnamefont {Gün}}, \ and\ \bibinfo {author} {\bibfnamefont {M.}~\bibnamefont {Hesse}},\ }\bibfield  {title} {\enquote {\bibinfo {title} {The parabolic anderson model on the hypercube},}\ }\href {\doibase https://doi.org/10.1016/j.spa.2019.09.016} {\bibfield  {journal} {\bibinfo  {journal} {Stoch. Process. Their Appl}\ }\textbf {\bibinfo {volume} {130}},\ \bibinfo {pages} {3369--3393} (\bibinfo {year} {2020})}\BibitemShut {NoStop}%
\bibitem [{\citenamefont {\v{S}tefa\v{n}\'ak}\ and\ \citenamefont {Skoup\'y}(2023)}]{Stefanak2023}%
  \BibitemOpen
  \bibfield  {author} {\bibinfo {author} {\bibfnamefont {M.}~\bibnamefont {\v{S}tefa\v{n}\'ak}}\ and\ \bibinfo {author} {\bibfnamefont {S.}~\bibnamefont {Skoup\'y}},\ }\bibfield  {title} {\enquote {\bibinfo {title} {Quantum walk state transfer on a hypercube},}\ }\href {\doibase 10.1088/1402-4896/acf3a2} {\bibfield  {journal} {\bibinfo  {journal} {Phys. Scr.}\ }\textbf {\bibinfo {volume} {98}},\ \bibinfo {pages} {104003} (\bibinfo {year} {2023})}\BibitemShut {NoStop}%
\bibitem [{\citenamefont {Laumann}\ \emph {et~al.}(2014)\citenamefont {Laumann}, \citenamefont {Pal},\ and\ \citenamefont {Scardicchio}}]{Laumann2014}%
  \BibitemOpen
  \bibfield  {author} {\bibinfo {author} {\bibfnamefont {C.~R.}\ \bibnamefont {Laumann}}, \bibinfo {author} {\bibfnamefont {A.}~\bibnamefont {Pal}}, \ and\ \bibinfo {author} {\bibfnamefont {A.}~\bibnamefont {Scardicchio}},\ }\bibfield  {title} {\enquote {\bibinfo {title} {Many-body mobility edge in a mean-field quantum spin glass},}\ }\href {\doibase 10.1103/PhysRevLett.113.200405} {\bibfield  {journal} {\bibinfo  {journal} {Phys. Rev. Lett.}\ }\textbf {\bibinfo {volume} {113}},\ \bibinfo {pages} {200405} (\bibinfo {year} {2014})}\BibitemShut {NoStop}%
\bibitem [{\citenamefont {Baldwin}\ \emph {et~al.}(2016)\citenamefont {Baldwin}, \citenamefont {Laumann}, \citenamefont {Pal},\ and\ \citenamefont {Scardicchio}}]{Baldwin2016}%
  \BibitemOpen
  \bibfield  {author} {\bibinfo {author} {\bibfnamefont {C.~L.}\ \bibnamefont {Baldwin}}, \bibinfo {author} {\bibfnamefont {C.~R.}\ \bibnamefont {Laumann}}, \bibinfo {author} {\bibfnamefont {A.}~\bibnamefont {Pal}}, \ and\ \bibinfo {author} {\bibfnamefont {A.}~\bibnamefont {Scardicchio}},\ }\bibfield  {title} {\enquote {\bibinfo {title} {The many-body localized phase of the quantum random energy model},}\ }\href {\doibase 10.1103/PhysRevB.93.024202} {\bibfield  {journal} {\bibinfo  {journal} {Phys. Rev. B}\ }\textbf {\bibinfo {volume} {93}},\ \bibinfo {pages} {024202} (\bibinfo {year} {2016})}\BibitemShut {NoStop}%
\bibitem [{\citenamefont {Roy}\ and\ \citenamefont {Logan}(2020)}]{Roy2020mbl}%
  \BibitemOpen
  \bibfield  {author} {\bibinfo {author} {\bibfnamefont {S.}~\bibnamefont {Roy}}\ and\ \bibinfo {author} {\bibfnamefont {D.~E.}\ \bibnamefont {Logan}},\ }\bibfield  {title} {\enquote {\bibinfo {title} {Fock-space correlations and the origins of many-body localization},}\ }\href {\doibase 10.1103/PhysRevB.101.134202} {\bibfield  {journal} {\bibinfo  {journal} {Phys. Rev. B}\ }\textbf {\bibinfo {volume} {101}},\ \bibinfo {pages} {134202} (\bibinfo {year} {2020})}\BibitemShut {NoStop}%
\bibitem [{\citenamefont {Logan}\ and\ \citenamefont {Welsh}(2019)}]{Logan2019}%
  \BibitemOpen
  \bibfield  {author} {\bibinfo {author} {\bibfnamefont {D.~E.}\ \bibnamefont {Logan}}\ and\ \bibinfo {author} {\bibfnamefont {S.}~\bibnamefont {Welsh}},\ }\bibfield  {title} {\enquote {\bibinfo {title} {Many-body localization in {F}ock space: {A} local perspective},}\ }\href {\doibase 10.1103/PhysRevB.99.045131} {\bibfield  {journal} {\bibinfo  {journal} {Phys. Rev. B}\ }\textbf {\bibinfo {volume} {99}},\ \bibinfo {pages} {045131} (\bibinfo {year} {2019})}\BibitemShut {NoStop}%
\bibitem [{\citenamefont {Scoquart}\ \emph {et~al.}(2024)\citenamefont {Scoquart}, \citenamefont {Gornyi},\ and\ \citenamefont {Mirlin}}]{Scoquart2024}%
  \BibitemOpen
  \bibfield  {author} {\bibinfo {author} {\bibfnamefont {T.}~\bibnamefont {Scoquart}}, \bibinfo {author} {\bibfnamefont {I.~V.}\ \bibnamefont {Gornyi}}, \ and\ \bibinfo {author} {\bibfnamefont {A.~D.}\ \bibnamefont {Mirlin}},\ }\bibfield  {title} {\enquote {\bibinfo {title} {Role of {F}ock-space correlations in many-body localization},}\ }\href {\doibase 10.1103/PhysRevB.109.214203} {\bibfield  {journal} {\bibinfo  {journal} {Phys. Rev. B}\ }\textbf {\bibinfo {volume} {109}},\ \bibinfo {pages} {214203} (\bibinfo {year} {2024})}\BibitemShut {NoStop}%
\bibitem [{\citenamefont {Regensburger}\ \emph {et~al.}(2011)\citenamefont {Regensburger}, \citenamefont {Bersch}, \citenamefont {Hinrichs}, \citenamefont {Onishchukov}, \citenamefont {Schreiber}, \citenamefont {Silberhorn},\ and\ \citenamefont {Peschel}}]{Regensburger2011}%
  \BibitemOpen
  \bibfield  {author} {\bibinfo {author} {\bibfnamefont {A.}~\bibnamefont {Regensburger}}, \bibinfo {author} {\bibfnamefont {C.}~\bibnamefont {Bersch}}, \bibinfo {author} {\bibfnamefont {B.}~\bibnamefont {Hinrichs}}, \bibinfo {author} {\bibfnamefont {G.}~\bibnamefont {Onishchukov}}, \bibinfo {author} {\bibfnamefont {A.}~\bibnamefont {Schreiber}}, \bibinfo {author} {\bibfnamefont {C.}~\bibnamefont {Silberhorn}}, \ and\ \bibinfo {author} {\bibfnamefont {U.}~\bibnamefont {Peschel}},\ }\bibfield  {title} {\enquote {\bibinfo {title} {Photon propagation in a discrete fiber network: {A}n interplay of coherence and losses},}\ }\href {\doibase 10.1103/PhysRevLett.107.233902} {\bibfield  {journal} {\bibinfo  {journal} {Phys. Rev. Lett.}\ }\textbf {\bibinfo {volume} {107}},\ \bibinfo {pages} {233902} (\bibinfo {year} {2011})}\BibitemShut {NoStop}%
\bibitem [{\citenamefont {Yuan}\ \emph {et~al.}(2018)\citenamefont {Yuan}, \citenamefont {Lin}, \citenamefont {Xiao},\ and\ \citenamefont {Fan}}]{Yuan2018}%
  \BibitemOpen
  \bibfield  {author} {\bibinfo {author} {\bibfnamefont {L.}~\bibnamefont {Yuan}}, \bibinfo {author} {\bibfnamefont {Q.}~\bibnamefont {Lin}}, \bibinfo {author} {\bibfnamefont {M.}~\bibnamefont {Xiao}}, \ and\ \bibinfo {author} {\bibfnamefont {S.}~\bibnamefont {Fan}},\ }\bibfield  {title} {\enquote {\bibinfo {title} {Synthetic dimension in photonics},}\ }\href {\doibase 10.1364/OPTICA.5.001396} {\bibfield  {journal} {\bibinfo  {journal} {Optica}\ }\textbf {\bibinfo {volume} {5}},\ \bibinfo {pages} {1396--1405} (\bibinfo {year} {2018})}\BibitemShut {NoStop}%
\bibitem [{\citenamefont {Hu}\ \emph {et~al.}(2020)\citenamefont {Hu}, \citenamefont {Reimer}, \citenamefont {Shams-Ansari}, \citenamefont {Zhang},\ and\ \citenamefont {Loncar}}]{Hu20}%
  \BibitemOpen
  \bibfield  {author} {\bibinfo {author} {\bibfnamefont {Y.}~\bibnamefont {Hu}}, \bibinfo {author} {\bibfnamefont {C.}~\bibnamefont {Reimer}}, \bibinfo {author} {\bibfnamefont {A.}~\bibnamefont {Shams-Ansari}}, \bibinfo {author} {\bibfnamefont {M.}~\bibnamefont {Zhang}}, \ and\ \bibinfo {author} {\bibfnamefont {M.}~\bibnamefont {Loncar}},\ }\bibfield  {title} {\enquote {\bibinfo {title} {Realization of high-dimensional frequency crystals in electro-optic microcombs},}\ }\href {\doibase 10.1364/OPTICA.395114} {\bibfield  {journal} {\bibinfo  {journal} {Optica}\ }\textbf {\bibinfo {volume} {7}},\ \bibinfo {pages} {1189--1194} (\bibinfo {year} {2020})}\BibitemShut {NoStop}%
\bibitem [{\citenamefont {Leefmans}\ \emph {et~al.}(2022)\citenamefont {Leefmans}, \citenamefont {Dutt}, \citenamefont {Williams}, \citenamefont {Yuan}, \citenamefont {Parto}, \citenamefont {Nori}, \citenamefont {Fan},\ and\ \citenamefont {Marandi}}]{Leefmans2022}%
  \BibitemOpen
  \bibfield  {author} {\bibinfo {author} {\bibfnamefont {C.}~\bibnamefont {Leefmans}}, \bibinfo {author} {\bibfnamefont {A.}~\bibnamefont {Dutt}}, \bibinfo {author} {\bibfnamefont {J.}~\bibnamefont {Williams}}, \bibinfo {author} {\bibfnamefont {L.}~\bibnamefont {Yuan}}, \bibinfo {author} {\bibfnamefont {M.}~\bibnamefont {Parto}}, \bibinfo {author} {\bibfnamefont {F.}~\bibnamefont {Nori}}, \bibinfo {author} {\bibfnamefont {S.}~\bibnamefont {Fan}}, \ and\ \bibinfo {author} {\bibfnamefont {A.}~\bibnamefont {Marandi}},\ }\bibfield  {title} {\enquote {\bibinfo {title} {Topological dissipation in a time-multiplexed photonic resonator network},}\ }\href {\doibase https://doi.org/10.1038/s41567-021-01492-w} {\bibfield  {journal} {\bibinfo  {journal} {Nat. Phys.}\ }\textbf {\bibinfo {volume} {18}},\ \bibinfo {pages} {442} (\bibinfo {year} {2022})}\BibitemShut {NoStop}%
\bibitem [{\citenamefont {Parto}\ \emph {et~al.}(2023)\citenamefont {Parto}, \citenamefont {Leefmans}, \citenamefont {Williams}, \citenamefont {Nori},\ and\ \citenamefont {Marandi}}]{Parto2023}%
  \BibitemOpen
  \bibfield  {author} {\bibinfo {author} {\bibfnamefont {M.}~\bibnamefont {Parto}}, \bibinfo {author} {\bibfnamefont {C.}~\bibnamefont {Leefmans}}, \bibinfo {author} {\bibfnamefont {J.}~\bibnamefont {Williams}}, \bibinfo {author} {\bibfnamefont {F.}~\bibnamefont {Nori}}, \ and\ \bibinfo {author} {\bibfnamefont {A.}~\bibnamefont {Marandi}},\ }\bibfield  {title} {\enquote {\bibinfo {title} {Non-{A}belian effects in dissipative photonic topological lattices},}\ }\href {https://www.nature.com/articles/s41467-023-37065-z} {\bibfield  {journal} {\bibinfo  {journal} {Nat. Commun.}\ }\textbf {\bibinfo {volume} {14}} (\bibinfo {year} {2023})}\BibitemShut {NoStop}%
\bibitem [{\citenamefont {Ehrhardt}\ \emph {et~al.}(2023)\citenamefont {Ehrhardt}, \citenamefont {Weidemann}, \citenamefont {Maczewsky}, \citenamefont {Heinrich},\ and\ \citenamefont {Szameit}}]{Ehrhardt2023}%
  \BibitemOpen
  \bibfield  {author} {\bibinfo {author} {\bibfnamefont {M.}~\bibnamefont {Ehrhardt}}, \bibinfo {author} {\bibfnamefont {S.}~\bibnamefont {Weidemann}}, \bibinfo {author} {\bibfnamefont {L.~J.}\ \bibnamefont {Maczewsky}}, \bibinfo {author} {\bibfnamefont {M.}~\bibnamefont {Heinrich}}, \ and\ \bibinfo {author} {\bibfnamefont {A.}~\bibnamefont {Szameit}},\ }\bibfield  {title} {\enquote {\bibinfo {title} {A perspective on synthetic dimensions in photonics},}\ }\href {\doibase https://doi.org/10.1002/lpor.202200518} {\bibfield  {journal} {\bibinfo  {journal} {Laser Photonics Rev.}\ }\textbf {\bibinfo {volume} {17}},\ \bibinfo {pages} {2200518} (\bibinfo {year} {2023})}\BibitemShut {NoStop}%
\bibitem [{\citenamefont {Leefmans}\ \emph {et~al.}(2024)\citenamefont {Leefmans}, \citenamefont {Parto}, \citenamefont {Williams} \emph {et~al.}}]{Leefmans2024}%
  \BibitemOpen
  \bibfield  {author} {\bibinfo {author} {\bibfnamefont {C.R.}\ \bibnamefont {Leefmans}}, \bibinfo {author} {\bibfnamefont {M.}~\bibnamefont {Parto}}, \bibinfo {author} {\bibfnamefont {J.}~\bibnamefont {Williams}},  \emph {et~al.},\ }\bibfield  {title} {\enquote {\bibinfo {title} {Topological temporally mode-locked laser},}\ }\href {\doibase 10.1038/s41567-024-02420-4} {\bibfield  {journal} {\bibinfo  {journal} {Nat. Phys.}\ }\textbf {\bibinfo {volume} {20}},\ \bibinfo {pages} {852} (\bibinfo {year} {2024})}\BibitemShut {NoStop}%
\bibitem [{\citenamefont {Coxeter}(1973)}]{CoxeterBook}%
  \BibitemOpen
  \bibfield  {author} {\bibinfo {author} {\bibfnamefont {H.~S.~M.}\ \bibnamefont {Coxeter}},\ }\href@noop {} {\emph {\bibinfo {title} {Regular Polytopes}}}\ (\bibinfo  {publisher} {Dover Publications; 3rd edition},\ \bibinfo {address} {New York},\ \bibinfo {year} {1973})\BibitemShut {NoStop}%
\bibitem [{\citenamefont {Arkhipov}\ \emph {et~al.}(2023)\citenamefont {Arkhipov}, \citenamefont {Miranowicz}, \citenamefont {Nori}, \citenamefont {\"Ozdemir},\ and\ \citenamefont {Minganti}}]{arkhipov2023c}%
  \BibitemOpen
  \bibfield  {author} {\bibinfo {author} {\bibfnamefont {I.~I.}\ \bibnamefont {Arkhipov}}, \bibinfo {author} {\bibfnamefont {A.}~\bibnamefont {Miranowicz}}, \bibinfo {author} {\bibfnamefont {F.}~\bibnamefont {Nori}}, \bibinfo {author} {\bibfnamefont {{\c{S}}.~K.}\ \bibnamefont {\"Ozdemir}}, \ and\ \bibinfo {author} {\bibfnamefont {F.}~\bibnamefont {Minganti}},\ }\bibfield  {title} {\enquote {\bibinfo {title} {Fully solvable finite simplex lattices with open boundaries in arbitrary dimensions},}\ }\href {\doibase 10.1103/PhysRevResearch.5.043092} {\bibfield  {journal} {\bibinfo  {journal} {Phys. Rev. Res.}\ }\textbf {\bibinfo {volume} {5}},\ \bibinfo {pages} {043092} (\bibinfo {year} {2023})}\BibitemShut {NoStop}%
\bibitem [{\citenamefont {Harary}\ \emph {et~al.}(1988)\citenamefont {Harary}, \citenamefont {Hayes},\ and\ \citenamefont {Wu}}]{Harary1988}%
  \BibitemOpen
  \bibfield  {author} {\bibinfo {author} {\bibfnamefont {F.}~\bibnamefont {Harary}}, \bibinfo {author} {\bibfnamefont {J.~P.}\ \bibnamefont {Hayes}}, \ and\ \bibinfo {author} {\bibfnamefont {H.-J.}\ \bibnamefont {Wu}},\ }\bibfield  {title} {\enquote {\bibinfo {title} {A survey of the theory of hypercube graphs},}\ }\href {\doibase https://doi.org/10.1016/0898-1221(88)90213-1} {\bibfield  {journal} {\bibinfo  {journal} {Comput. Math. Appl.}\ }\textbf {\bibinfo {volume} {15}},\ \bibinfo {pages} {277--289} (\bibinfo {year} {1988})}\BibitemShut {NoStop}%
\bibitem [{\citenamefont {Maldacena}\ and\ \citenamefont {Qi}(2018)}]{maldacena2018}%
  \BibitemOpen
  \bibfield  {author} {\bibinfo {author} {\bibfnamefont {J.}~\bibnamefont {Maldacena}}\ and\ \bibinfo {author} {\bibfnamefont {X.-L.}\ \bibnamefont {Qi}},\ }\href {https://arxiv.org/abs/1804.00491} {\enquote {\bibinfo {title} {Eternal traversable wormhole},}\ } (\bibinfo {year} {2018}),\ \Eprint {http://arxiv.org/abs/1804.00491} {arXiv:1804.00491 [hep-th]} \BibitemShut {NoStop}%
\bibitem [{\citenamefont {Chiu}\ \emph {et~al.}(2016)\citenamefont {Chiu}, \citenamefont {Teo}, \citenamefont {Schnyder},\ and\ \citenamefont {Ryu}}]{Chiu2016}%
  \BibitemOpen
  \bibfield  {author} {\bibinfo {author} {\bibfnamefont {C.-K.}\ \bibnamefont {Chiu}}, \bibinfo {author} {\bibfnamefont {J.~C.~Y.}\ \bibnamefont {Teo}}, \bibinfo {author} {\bibfnamefont {A.~P.}\ \bibnamefont {Schnyder}}, \ and\ \bibinfo {author} {\bibfnamefont {S.}~\bibnamefont {Ryu}},\ }\bibfield  {title} {\enquote {\bibinfo {title} {Classification of topological quantum matter with symmetries},}\ }\href {\doibase 10.1103/RevModPhys.88.035005} {\bibfield  {journal} {\bibinfo  {journal} {Rev. Mod. Phys.}\ }\textbf {\bibinfo {volume} {88}},\ \bibinfo {pages} {035005} (\bibinfo {year} {2016})}\BibitemShut {NoStop}%
\bibitem [{\citenamefont {Ramachandran}\ \emph {et~al.}(2017)\citenamefont {Ramachandran}, \citenamefont {Andreanov},\ and\ \citenamefont {Flach}}]{Ramachandran2017}%
  \BibitemOpen
  \bibfield  {author} {\bibinfo {author} {\bibfnamefont {A.}~\bibnamefont {Ramachandran}}, \bibinfo {author} {\bibfnamefont {A.}~\bibnamefont {Andreanov}}, \ and\ \bibinfo {author} {\bibfnamefont {S.}~\bibnamefont {Flach}},\ }\bibfield  {title} {\enquote {\bibinfo {title} {Chiral flat bands: {E}xistence, engineering, and stability},}\ }\href {\doibase 10.1103/PhysRevB.96.161104} {\bibfield  {journal} {\bibinfo  {journal} {Phys. Rev. B}\ }\textbf {\bibinfo {volume} {96}},\ \bibinfo {pages} {161104} (\bibinfo {year} {2017})}\BibitemShut {NoStop}%
\bibitem [{\citenamefont {Nelson}\ and\ \citenamefont {Widom}(1984)}]{Nelson1984}%
  \BibitemOpen
  \bibfield  {author} {\bibinfo {author} {\bibfnamefont {D.~R.}\ \bibnamefont {Nelson}}\ and\ \bibinfo {author} {\bibfnamefont {M.}~\bibnamefont {Widom}},\ }\bibfield  {title} {\enquote {\bibinfo {title} {Symmetry, {L}andau theory and polytope models of glass},}\ }\href {\doibase https://doi.org/10.1016/0550-3213(84)90281-5} {\bibfield  {journal} {\bibinfo  {journal} {Nucl. Phys. B}\ }\textbf {\bibinfo {volume} {240}},\ \bibinfo {pages} {113--139} (\bibinfo {year} {1984})}\BibitemShut {NoStop}%
\bibitem [{Note1()}]{Note1}%
  \BibitemOpen
  \bibinfo {note} {Indeed, each matrix $S_k$ in \protect \mbox {Eq.~(\ref {An})} with $\alpha _k=\beta _k=0$ has two eigenstates $\psi _1=[1,1]^T$ and $\psi _2=[-1,1]^T$, corresponding to eigenvalues $\lambda _{1,2}=\pm 1$. As such, the $2n!/[n!]^2$-folded degenerate ZESs for a given Hamiltonian $H_{2n}$, are formed by $2n$-folded Kronecker products of $n!$ number of states $\psi _1$ and the same number of states $\psi _2$.}\BibitemShut {Stop}%
\bibitem [{\citenamefont {Karle}\ \emph {et~al.}(2021)\citenamefont {Karle}, \citenamefont {Serbyn},\ and\ \citenamefont {Michailidis}}]{Karle2021}%
  \BibitemOpen
  \bibfield  {author} {\bibinfo {author} {\bibfnamefont {V.}~\bibnamefont {Karle}}, \bibinfo {author} {\bibfnamefont {M.}~\bibnamefont {Serbyn}}, \ and\ \bibinfo {author} {\bibfnamefont {A.~A.}\ \bibnamefont {Michailidis}},\ }\bibfield  {title} {\enquote {\bibinfo {title} {Area-law entangled eigenstates from nullspaces of local {H}amiltonians},}\ }\href {\doibase 10.1103/PhysRevLett.127.060602} {\bibfield  {journal} {\bibinfo  {journal} {Phys. Rev. Lett.}\ }\textbf {\bibinfo {volume} {127}},\ \bibinfo {pages} {060602} (\bibinfo {year} {2021})}\BibitemShut {NoStop}%
\bibitem [{\citenamefont {Turner}\ \emph {et~al.}(2018)\citenamefont {Turner}, \citenamefont {Michailidis}, \citenamefont {Abanin}, \citenamefont {Serbyn},\ and\ \citenamefont {Papi\ifmmode~\acute{c}\else \'{c}\fi{}}}]{Turner2018}%
  \BibitemOpen
  \bibfield  {author} {\bibinfo {author} {\bibfnamefont {C.~J.}\ \bibnamefont {Turner}}, \bibinfo {author} {\bibfnamefont {A.~A.}\ \bibnamefont {Michailidis}}, \bibinfo {author} {\bibfnamefont {D.~A.}\ \bibnamefont {Abanin}}, \bibinfo {author} {\bibfnamefont {M.}~\bibnamefont {Serbyn}}, \ and\ \bibinfo {author} {\bibfnamefont {Z.}~\bibnamefont {Papi\ifmmode~\acute{c}\else \'{c}\fi{}}},\ }\bibfield  {title} {\enquote {\bibinfo {title} {Quantum scarred eigenstates in a {R}ydberg atom chain: {E}ntanglement, breakdown of thermalization, and stability to perturbations},}\ }\href {\doibase 10.1103/PhysRevB.98.155134} {\bibfield  {journal} {\bibinfo  {journal} {Phys. Rev. B}\ }\textbf {\bibinfo {volume} {98}},\ \bibinfo {pages} {155134} (\bibinfo {year} {2018})}\BibitemShut {NoStop}%
\bibitem [{\citenamefont {Schecter}\ and\ \citenamefont {Iadecola}(2018)}]{Schecter2018}%
  \BibitemOpen
  \bibfield  {author} {\bibinfo {author} {\bibfnamefont {M.}~\bibnamefont {Schecter}}\ and\ \bibinfo {author} {\bibfnamefont {T.}~\bibnamefont {Iadecola}},\ }\bibfield  {title} {\enquote {\bibinfo {title} {Many-body spectral reflection symmetry and protected infinite-temperature degeneracy},}\ }\href {\doibase 10.1103/PhysRevB.98.035139} {\bibfield  {journal} {\bibinfo  {journal} {Phys. Rev. B}\ }\textbf {\bibinfo {volume} {98}},\ \bibinfo {pages} {035139} (\bibinfo {year} {2018})}\BibitemShut {NoStop}%
\bibitem [{Note2()}]{Note2}%
  \BibitemOpen
  \bibinfo {note} {The other eigenvector with the eigenvalue $\lambda _k=\alpha _k+\alpha _k^{-1}$ attains the form $\psi _{k,\lambda \protect \neq 0}\equiv \protect \Big [\alpha _k, 1\protect \Big ]^T$.}\BibitemShut {Stop}%
\bibitem [{\citenamefont {Choi}\ \emph {et~al.}(2016)\citenamefont {Choi}, \citenamefont {Hild}, \citenamefont {Zeiher} \emph {et~al.}}]{Choi2016}%
  \BibitemOpen
  \bibfield  {author} {\bibinfo {author} {\bibfnamefont {J.-y.}\ \bibnamefont {Choi}}, \bibinfo {author} {\bibfnamefont {S.}~\bibnamefont {Hild}}, \bibinfo {author} {\bibfnamefont {J.}~\bibnamefont {Zeiher}},  \emph {et~al.},\ }\bibfield  {title} {\enquote {\bibinfo {title} {Exploring the many-body localization transition in two dimensions},}\ }\href {\doibase 10.1126/science.aaf8834} {\bibfield  {journal} {\bibinfo  {journal} {Science}\ }\textbf {\bibinfo {volume} {352}},\ \bibinfo {pages} {1547–1552} (\bibinfo {year} {2016})}\BibitemShut {NoStop}%
\bibitem [{\citenamefont {Pastur}(1980)}]{Pastur1980}%
  \BibitemOpen
  \bibfield  {author} {\bibinfo {author} {\bibfnamefont {L.~A.}\ \bibnamefont {Pastur}},\ }\bibfield  {title} {\enquote {\bibinfo {title} {Spectral properties of disordered systems in the one-body approximation},}\ }\href {\doibase 10.1007/BF01222516} {\bibfield  {journal} {\bibinfo  {journal} {Commun. Math. Phys.}\ }\textbf {\bibinfo {volume} {75}},\ \bibinfo {pages} {179--196} (\bibinfo {year} {1980})}\BibitemShut {NoStop}%
\end{thebibliography}%
\end{document}